\documentclass[twocolumn]{aastex62}

\submitjournal{ApJ}
\usepackage[all]{hypcap} 
\usepackage{tabularx, graphicx}
\usepackage{gensymb}

\usepackage{bm}
\usepackage{amsmath}
\usepackage{color}

\newcommand{\Msun}{\,M_{\odot}}

\newcommand{\dls}{D_\mathrm{LS}}
\newcommand{\ds}{D_\mathrm{S}}
\newcommand{\dl}{D_\mathrm{L}}
\newcommand{\zs}{z_\mathrm{S}}
\newcommand{\zl}{z_\mathrm{L}}

\newcommand{\vect}[1]{\bm {#1}}
\newcommand{\ML}{M(<\theta_E)}
\newcommand{\MT}{M_{sim}(<\theta_E)}

\newcommand{\FXAllScatter}{$13.9$\%}
\newcommand{\FXAllBias}{$8.8$\%}
\newcommand{\FRAllScatter}{$27.4$\%}
\newcommand{\FRAllBias}{$20.2$\%}
\newcommand{\FXAllBCGScatter}{$14.8$\%}
\newcommand{\FXAllBCGBias}{$10.2$\%}
\newcommand{\FXNoZScatter}{$13.8$\%}
\newcommand{\FXNoZbias}{$9.0$\%}
\newcommand{\FXNoZBCGScatter}{$18.2$\%}
\newcommand{\FXNoZBCGbias}{$8.5$\%}
\newcommand{\FXECScatter}{$10.1$\%}
\newcommand{\FXECBias}{$-0.4$\%}
\newcommand{\FXBCGECScatter}{$10.9$\%}
\newcommand{\FXBCGECBias}{$-0.3$\%}

\newcommand{\KSStat}{{$0.0896$}}
\newcommand{\KSPvalue}{{$0.0402$}}

\begin{document}
\title{Efficient Mass Estimate at the Core of Strong Lensing Galaxy Clusters Using the Einstein Radius}

\author[0000-0002-7868-9827]{J. D. Remolina Gonz\'{a}lez}
\affiliation{Department of Astronomy, University of Michigan, 1085 S. University Ave, Ann Arbor, MI 48109, USA}
\email{jremolin@umich.edu}

\author[0000-0002-7559-0864]{K. Sharon}
\affiliation{Department of Astronomy, University of Michigan, 1085 S. University Ave, Ann Arbor, MI 48109, USA}

\author[0000-0002-7775-5423]{B. Reed}
\affil{Department of Astronomy, University of Michigan, 1085 S. University Ave, Ann Arbor, MI 48109, USA}
\affil{Department of Astronomy, Indiana University, Bloomington, IN 47405, USA}

\author[0000-0001-6800-7389]{N. Li}
\affil{CAS, Key Laboratory of Space Astronomy and Technology, National Astronomical Observatories, A20 Datun Road, Chaoyang District, Beijing 100012, People’s Republic of China}
\affil{School of Physics and Astronomy, Nottingham University, University Park, Nottingham NG7 2RD, UK}

\author[0000-0003-3266-2001]{G. Mahler}
\affil{Department of Astronomy, University of Michigan, 1085 S. University Ave, Ann Arbor, MI 48109, USA}

\author[0000-0001-7665-5079]{L. E. Bleem}
\affil{Argonne National Laboratory, High-Energy Physics Division, Argonne, IL 60439}
\affil{Kavli Institute for Cosmological Physics, University of Chicago, 5640 South Ellis Avenue, Chicago, IL 60637, USA}

\author[0000-0003-1370-5010]{M. Gladders}
\affil{Department of Astronomy and Astrophysics, University of Chicago, 5640 South Ellis Avenue, Chicago, IL 60637, USA}
\affil{Kavli Institute for Cosmological Physics, University of Chicago, 5640 South Ellis Avenue, Chicago, IL 60637, USA}

\author[0000-0003-3791-2647]{A. Niemiec}
\affil{Department of Astronomy, University of Michigan, 1085 S. University Ave, Ann Arbor, MI 48109, USA}
\affil{Centre for Extragalactic Astronomy, Department of Physics, Durham University, Durham DH1 3LE, UK}
\affil{Institute for Computational Cosmology, Durham University, South Road, Durham DH1 3LE, UK}

\author[0000-0003-3108-9039]{A. Acebron}
\affil{Physics Department, Ben-Gurion University of the Negev, P.O. Box 653, Be'er-Sheva 8410501, Israel}

\author[0000-0002-6471-5369]{H. Child}
\affil{Argonne National Laboratory, High-Energy Physics Division, Argonne, IL 60439}
\affil{Department of Astronomy and Astrophysics, University of Chicago, 5640 South Ellis Avenue, Chicago, IL 60637, USA}

\begin{abstract}

In the era of large surveys, yielding thousands of galaxy clusters, efficient mass proxies at all scales are necessary in order to fully utilize clusters as cosmological probes. At the cores of strong lensing clusters, the Einstein radius can be turned into a mass estimate. This efficient method has been routinely used in literature, in lieu of detailed mass models; however, its scatter, assumed to be $\sim30\%$, has not yet been quantified. Here, we assess this method by testing it against ray-traced images of cluster-scale halos from the Outer Rim N-body cosmological simulation. We measure a scatter of \FXAllScatter\ and a positive bias of \FXAllBias\ in $\ML$, with no systematic correlation with total cluster mass, concentration, or lens or source redshifts.  We find that increased deviation from spherical symmetry increases the scatter; conversely, where the lens produces arcs that cover a large fraction of its Einstein circle, both the scatter and the bias decrease.  While spectroscopic redshifts of the lensed sources are critical for accurate magnifications and time delays, we show that for the purpose of estimating the total enclosed mass, the scatter introduced by source redshift uncertainty is negligible compared to other sources of error. Finally, we derive and apply an empirical correction that eliminates the bias, and reduces the scatter to \FXECScatter\ without introducing new correlations with mass, redshifts, or concentration. Our analysis provides the first quantitative assessment of the uncertainties in $\ML$, and enables its effective use as a core mass estimator of strong lensing galaxy clusters. 

\end{abstract}



\section{Introduction} 
\label{sec:intro}

Located at the knots of the cosmic web, galaxy clusters trace regions of over-density in the large-scale structure of the universe, making them ideal cosmic laboratories. As cosmological probes (see review articles \citealt{Allen:11, Mantz:14}), clusters have been used to study dark energy (e.g., \citealt{Frieman:08, Heneka:18, Bonilla:18, Huterer:18}) and dark matter (e.g., \citealt{Bradac:06, Clowe:06, Bradac:08, Diego:18}), constrain cosmological parameters (e.g., \citealt{Gladders:07, Dunkley:09, Rozo:10, Mantz:10, Mantz:14, deHaan:16, Bocquet:19}), measure the baryonic fraction (e.g., \citealt{Fabian:91, Allen:08, Vikhlinin:09}) and the amplitude of the matter power spectrum (e.g., \citealt{Allen:03}). Crucial to cosmological studies using galaxy clusters is a large well-defined sample with a complete characterization of the selection function of the observations (e.g., \citealt{Hu:03b, Khedekar:13}).  

The mass distribution of galaxy clusters (cluster mass function) provides a connection between the observables and the underlying cosmology, and can constrain structure formation models (e.g., \citealt{Jenkins:01, Evrard:02, Corless:09}). The galaxy cluster dynamical and non-linear hierarchical merging growth process \citep{Bertschinger:98} introduces variance in the astronomical measurements \citep{Evrard:02, Allen:11, Huterer:18}. Understanding the systematic errors and assumptions made when estimating the mass of galaxy clusters is paramount as they depend on observable astrophysical quantities (e.g., \citealt{Evrard:02, Huterer:18}).

With the advent of recent and upcoming large surveys spanning a broad wavelength range, thousands of strong lensing galaxy clusters will be detected out to redshift of $z\sim 2$ with a high completeness and purity. Examples include the surveys from the South Pole Telescope (SPT-3G, \citealt{Benson:14}; SPT-SZ 2500 deg$^2$, \citealt{Bleem:15}), Atacama Cosmological Telescope (ACT, \citealt{Marriage:11,Hilton:18}), Cerro Chajnantor Atacama Telescope (CCAT, \citealt{Mittal:18}), Dark Energy Survey (DES, \citealt{Abbott:18}), Euclid \citep{Laureijs:11,Boldrin:12}, Vera Rubin Observatory Legacy Survey of Space and Time (LSST, \citealt{LSST:09}), ROSAT All-Sky Survey (RASS, \citealt{Ebeling:98,Ebeling:00}), and eROSITA \citep{Pillepich:18}. A thorough characterization of the selection function and bias implicit in the observations and detections is key. In addition, multi-wavelength coverage of some galaxy clusters will allow for an extensive study of their physical components.

Studies of the mass profile of galaxy clusters can provide us with information related to evolution of structure, formation and feedback processes, and dark matter properties. The methods used to estimate the mass of galaxy clusters include X-ray (e.g., \citealt{Vikhlinin:09, Ettori:19, Mantz:18}), Sunyaev-Zel'dovich effect (SZ, \citealt{Sunyaev:72, Sunyaev:80}; e.g., \citealt{Reichardt:13, Sifon:13, Planck:16}), richness (e.g., \citealt{Yee:03, Koester:07, Rykoff:16}), dynamics (e.g., \citealt{Gifford:13, Foex:17}), and gravitational lensing (e.g., \citealt{Kneib:11, Hoekstra:13, Sharon:15, Sharon:20}). Gravitational lensing (weak and strong) is the best technique to probe the total projected (baryonic and dark matter) mass density, independent of assumptions on the dynamical state of the cluster or baryonic physics. At the cores of galaxy clusters, strong gravitational lensing measures mass at the smallest radial scales and most extreme over-densities; when coupled with a mass proxy at a large radii, strong lensing can constrain global properties of the mass profile, including the concentration parameter. 

Advances in strong lens (SL) modeling, including better understanding of SL systematics \citep{Johnson:16},  its effects on constraining cosmological parameters \citep{Acebron:17}, magnification \citep{Priewe:17,Raney:20}, consequences due to the number of constraints \citep{Mahler:18},  and the use of spectroscopic and photometric redshifts \citep{Cerny:18}, make strong lens modeling a robust technique to study galaxy clusters and the background universe they magnify. A detailed lens model requires extensive follow-up:  (1) imaging to identify multiple images and (2) spectroscopy of the lensed images to obtain spectroscopic redshifts of the sources (e.g., \citealt{Johnson:14, Zitrin:14, Diego:16, Kawamata:16, Lotz:17, Strait:18, Lagattuta:19, Sebesta:19, Sharon:20}). The location of the multiple images and the spectroscopic redshifts are used as constraints when computing the SL models. Typically, a detailed SL model for a rich galaxy cluster can take weeks to finalize, and it is not an automated process. Given the large numbers of strong lensing galaxy clusters expected from coming surveys, an accurate, fast, and well-characterized method of extracting basic strong lensing information is needed.

In this paper, we evaluate the use of the geometric Einstein radius to estimate the mass at the core of SL galaxy clusters. We determine the uncertainties in the mass estimate, identify its limitations, investigate dependencies on the shape of the projected mass distribution, and find a possible empirical correction to de-bias the mass estimate. We base our analyses on the state-of-the-art, dark matter only, `Outer Rim' simulation \citep{Heitmann:19}. The Outer Rim contains a large sample of massive dark matter halos, and has sufficient mass resolution to enable precise and accurate ray-tracing of the strong lensing due to these halos.

This paper is organized as follows. In \S \ref{sec:lensing}, we describe the lensing formalism, including a detailed description of the assumptions of the Einstein radius method to compute the enclosed mass. In \S \ref{sec:sim}, we summarize the properties of the `Outer Rim' simulation, the halo sample used in our analysis, and the cosmological framework. In \S\ref{sec:methods}, we detail how we measure the Einstein radius from the ray-traced images and compute both the inferred mass enclosed by the Einstein radius and the true mass from the simulation. In \S \ref{sec:analysis}, we present our analysis of the mass estimate and the systematics that contribute to the scatter and bias. In \S \ref{sec:no_zs}, we investigate the effect of not having the redshift information of the background sources ($\zs$) on the mass estimate. In \S \ref{sec:emp_cor}, we propose an empirical correction to de-bias the mass estimate. Lastly, we present our conclusions and offer a prescription for applying our findings to real data in \S \ref{sec:conclusion}. 

For consistency with the simulations, we adopt a \textit{WMAP}-7 \citep{Komatsu:11} Flat $\Lambda$CDM cosmology in our analysis $\Omega_{\Lambda} = 0.735$, $\Omega_{M} = 0.265$, and $h = 0.71$. The large scale masses are reported in terms of M$_{\mathrm{Nc}}$, where M$_{\mathrm{Nc}}$ is defined as the mass enclosed within a radius at which the average density is N times the critical density of the universe at the cluster redshift. 


\section{BACKGROUND: Strong Gravitational Lensing}
\label{sec:lensing}

Gravitational lensing (see \citealt{Schneider:06a, Kneib:11} for reviews about gravitational lensing) occurs when photons deviate from their original direction as they travel to the observer through a locally curved space-time near a massive object, as described by Einstein's General Theory of Relativity. The lensing equation (\ref{eq:lenseq}) traces the image-plane position of images of lensed sources to the source plane location of the background sources. When multiple solutions to the lensing equation exist, multiply-imaged systems are possible, defining the strong lensing regime. The lensing equation is written as:

\begin{equation}
\begin{split}
	\vect{\beta} & = \vect{\theta} - \vect{\alpha} (\vect{\theta}), \\
    \vect{\alpha}(\vect{\theta}) & = \frac{\dls (\zl,\zs)}{\ds (\zs)} \vect{\hat{\alpha}} (\vect{\theta}),
\label{eq:lenseq}
\end{split}
\end{equation}

\noindent where $\vect{\beta}$ is the position of the lensed source in the source plane, $\vect{\theta}$ is the image plane location of the images, $\vect{\alpha}(\vect{\theta})$ is the deflection angle, $\dls (\zl,\zs)$ is the angular diameter distance between the lens and the source, $\ds (\zs)$ is the angular diameter distance between the observer and the source, $\zl$ is the redshift of the lens (in our case the redshift of the galaxy cluster), and $\zs$ is the redshift of the background source. The deflection angle depends on the gravitational potential of the cluster projected along the line-of-sight. 

The magnification, $\mu$, of a gravitational lens can be expressed as the determinant of the magnification matrix:

\begin{eqnarray}
\label{eq:mag}
	\mu^{-1} = det(\mathcal{A}^{-1})=(1-\kappa)^2-\gamma^2,
\end{eqnarray}

\noindent where $\kappa$ is the convergence and $\gamma$ is the shear. The locations of theoretical infinite magnification in the image plane are called the tangential and radial critical curves, naming the primary direction along which images (arcs) are magnified.

For a circularly symmetric lens with the origin centered at the point of symmetry, the angles $\vect{\alpha}(\vect{\theta})$ and $\vect{\beta}$ are collinear with $\vect{\theta}$. Then the lens equation (eq.~\ref{eq:lenseq}) becomes one-dimensional, $\beta = \theta - \alpha(\theta)$, and the deflection angle is:

\begin{equation}
\label{eq:deflec_angle}
\begin{split}
    \alpha(\theta) & = \frac{2}{\theta} \int_{0}^{\theta} \theta d\theta \kappa(\theta) \\
    & = \frac{4GM(<\theta)}{c^2 \theta} \frac{\dls (\zl,\zs)}{\dl (\zl) \ds (\zs)} \\
    & = \langle \kappa(\theta) \rangle \theta, 
\end{split}
\end{equation}

\noindent where $\dl (\zl)$ is the angular diameter distance from the observer to the lens, $c$ is the speed of light, and $G$ is the gravitational constant. We can then substitute the deflection angle into the one-dimensional lens equation:

\begin{equation}
\label{eq:1d_lenseq}
    \beta = \theta (1 - \langle \kappa (\theta) \rangle ),
\end{equation}

\noindent where the critical region, defined as $\langle \kappa(\theta)\rangle = 1$, defines the tangential critical curve. In this circularly symmetric case, $\alpha(\theta) = \theta$, \autoref{eq:deflec_angle} becomes

\begin{equation}
    \theta^2 = \frac{4GM(<\theta)}{c^2} \frac{\dls (\zl,\zs)}{\dl (\zl) \ds (\zs)}.
\end{equation}

\noindent Last, substituting the critical surface density, $\Sigma_{cr}(\zl,zs)$,

\begin{equation}
\label{eq:s_crit}
	\Sigma_{cr}(\zl,\zs) = \frac{c^2}{4\pi G}\frac{\ds(\zs)}{\dl(\zl) \dls(\zl,\zs)},
\end{equation}

\noindent we obtain the expression of the Einstein radius \citep{Narayan:96,Schneider:06a,Kochanek:06,Bartelmann:10,Kneib:11}:

\begin{equation}
\label{eq:er}
    \theta_E^2 = \frac{M(<\theta_E)}{\pi \Sigma_{cr}(\zl,\zs) \dl^2 (\zl)}.
\end{equation}

\noindent Re-arranging \autoref{eq:er}, the total projected mass enclosed by the Einstein radius  of a circularly symmetric lens can be computed as:

\begin{eqnarray}
\label{eq:m_er}
	M(<\theta_E)=\Sigma_{cr} (\zl,\zs) \  \pi \ [\dl(\zl)\theta_E]^2.
\end{eqnarray}

An Einstein ring results from the exact alignment of the source, lens, and observer, as well as the circular symmetry of the lens. This causes an observed  ring-like feature to appear around the lens. However, the three-dimensional mass density distribution of both simulated halos and observed clusters is better described by a triaxial ellipsoid \citep{Wang:2009, Despali:14, Bonamigo:15}. Complete Einstein rings are not often observed around clusters due to the more complex mass distribution; nevertheless, authors often use the clustercentric projected distance to a giant arc as a proxy for the Einstein radius. The mass calculated using \autoref{eq:m_er} is useful for the study of galaxy clusters, since it provides a quick estimate of the mass within the Einstein radius. It was estimated to produce a scatter of $\sim 30\%$  with respect to the true mass enclosed \citep{Bartelmann:96,Schneider:06b}. This uncertainty was adopted in the literature extensively when estimating the total projected mass enclosed by the Einstein radius (e.g., \citealt{Allam:07, Belokurov:07, Werner:07, Diehl:09, Bettinelli:16, Dahle:16, Nord:16}), despite limited quantification of its accuracy and precision. 


\section{DATA: Simulated Lenses} 
\label{sec:sim}

\subsection{The Outer Rim Simulation}
\label{subsec:outer_rim}

To assess the accuracy and precision of the enclosed mass inferred from the Einstein radius, we use the state-of-the-art, large-volume, high-mass-resolution, gravity-only, N-body simulation `Outer Rim' \citep{Heitmann:19} with the HACC framework \citep{Habib:16} carried out at the Blue Gene/Q (BG/Q) system Mira at Argonne National Laboratory. The cosmology used assumes a Flat-$\Lambda$CDM model, with parameters adopted from WMAP-7 \citep{Komatsu:11}, $\mathrm{h} = 0.71$ and $\Omega_M = 0.264789$. The size of the simulation box on the side is $L = 3000 \ \mathrm{Mpc} \ \mathrm{h}^{-1}$ and it evolves $10,240^3 \approx 1.1$ trillion particles with a mass resolution of $m_p = 1.85\times \ 10^9 \ \mathrm{\Msun} \ \mathrm{h}^{-1}$ and a force resolution in co-moving units of $3 \ \mathrm{kpc} \ \mathrm{h}^{-1}$.

The large volume of the simulation run allows for many massive halos to be included in the same simulation box, covering the redshift range of interest ($z \sim 0.1 - 0.7$), and the high mass resolution provides excellent projected mass profile distributions of the individual clusters. The large number of massive halos allows for a rigorous statistical analysis, representative of the universe and is sufficient to enable strong lensing computations without the need of re-simulation. In previous simulation efforts when small numbers of massive halos were present in the simulation box, re-simulation of those halos was done to increase the sample to better the statistics \citep{Meneghetti:08, Meneghetti:10}. The Outer Rim, amongst other applications, was used to study dark matter halo profiles and the concentration-Mass relation \citep{Child:18} and to construct realistic strong lensing simulated images \citep{Li:16}.  

The majority of the mass in galaxy clusters is in the form of dark matter. Baryons contribute mostly at the core of the galaxy cluster, where the brightest cluster galaxy (BCG) and the hot intra-cluster medium (ICM) reside. Studies have found non-negligible baryonic effects from subhaloes of satellite galaxies as well as the BCG at small $\theta_E$ scale \citep{Meneghetti:03, Wambsganss:04, Oguri:06, Hilbert:07, Hilbert:08, Wambsganss:08, Oguri:09}. Fully accounting for these baryonic effects awaits for simulations that include baryonic physics in large cosmological boxes.

\subsection{Simulated SPT-like Strong Lensing Sample}
\label{subsec:sl_sample}

Galaxy cluster halos were identified in the simulation using a friends-of-friends algorithm with a unit-less linking length of $b=0.168$ \citep{Heitmann:19}. The surface mass density was then computed using a density estimator. Extensive testing by \citet{Rangel:16} showed that the mass resolution is robust enough to compute strong lensing for halos with masses $\mathrm{M}_{500} > 2 \times 10^{14} \ \mathrm{\Msun} \ \mathrm{h}^{-1}$. Following an SPT-like selection function, the halos with a mass larger than $\mathrm{M}_{500} > 2.1 \times 10^{14} \ \mathrm{\Msun} \ \mathrm{h}^{-1}$ were selected to form the cluster sample.

The simulated halo masses ($M_{500}$, $M_{200}$) and concentrations ($c_{200}$) that we use in this work were calculated by \citet{Li:19} and \citet{Child:18}. We adopt the dynamical state values and definitions from \citet{Child:18}; a dynamically-relaxed cluster is identified where the distance between the dark matter halo center and the spherical over-density center is smaller than $0.7 R_{200}$. When referring to the dynamical state of the galaxy cluster, the center was defined as the center of the potential from all the particles in the simulation corresponding to the particular dark matter halo.

To select SL clusters out of the mass-limited sample, we first compute   $\kappa(\theta)$ for a background source redshift of $z=2$ for each line of sight.
We then identify strong lensing clusters as all lines of sight for which the Einstein radius of the critical region that satisfies $\langle \kappa(\theta) \rangle = 1$ is larger than a few arcseconds. The resulting sample of SPT-like simulated strong lenses includes $74$ galaxy cluster halos spanning the redshift range of $z_L \sim 0.16 - 0.67$.

\begin{figure*}
\center
\includegraphics[width=1\textwidth]{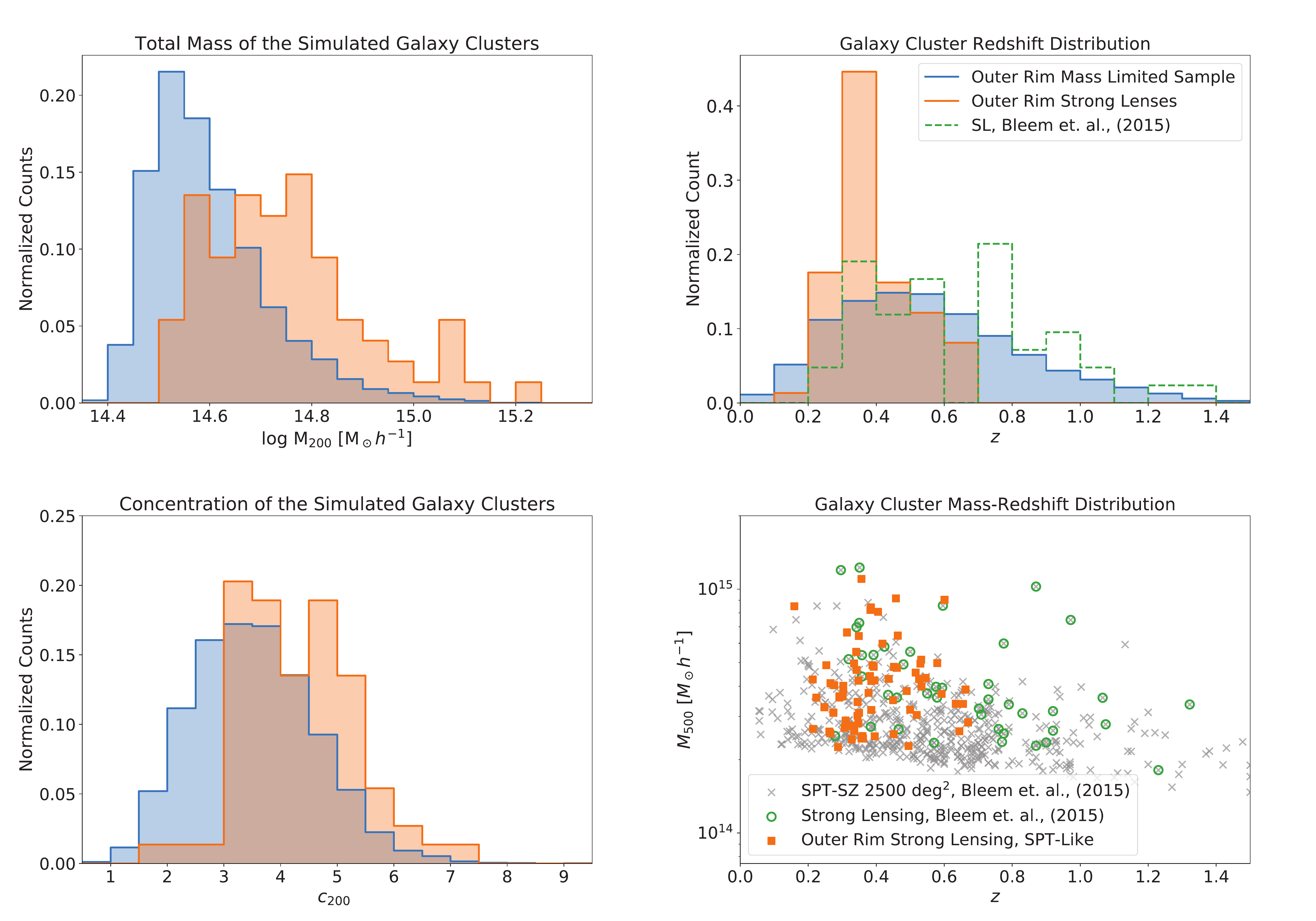}
\caption{\textsc{\textbf{Properties of the Simulated Sample.}} 
\textit{Top-Left}: the total mass ($M_{200}$), \textit{Top-Right}: redshift ($z$), and \textit{Bottom-Left}: concentration ($c_{200}$) distributions of the simulated halos. The mass-limited sample is shown in blue, and strong lenses are in orange. The masses and concentrations were computed by \citet{Li:19} and \citet{Child:18}. The counts are normalized by the total number of halos in each sample.
\textit{Bottom-Right}: the mass-redshift distribution ($M_{500}$ - $z$). Orange squares indicate the Outer Rim strong lensing cluster halos; grey crosses are observed clusters from the 2500-Square-Degree SPT-SZ Survey (\citealt{Bleem:15}). The green circles, and the green dotted line in the \textit{Right} panels, are strong lensing galaxy clusters from \citet{Bleem:15}, which were identified from very heterogeneous imaging data and are likely not representative of all the strong lenses in the SPT sample.}
\label{fig:cluster_sample}
\end{figure*}

In \autoref{fig:cluster_sample}, we summarize some of the halo properties of the mass-limited sample and the SL sample. The first three panels show the distributions of redshifts, masses, and concentrations. As can be seen in these panels,  the distribution of strong lensing clusters peaks at higher total mass, higher concentration, and lower redshift than the mass-limited sample. Similar trends have been identified in both simulations \citep{Oguri:11,Giocoli:14} and observations \citep{Gralla:11,Oguri:12}.

In the forth panel,  we plot the mass-redshift distribution of the simulated clusters and that of the observed clusters from the SPT-SZ 2500 deg$^2$ survey \citep{Bleem:15}.

As can be seen in the right panels of \autoref{fig:cluster_sample}, the \citet{Bleem:15} strong lensing sample extends to higher cluster redshifts than our simulated sample.
The effective redshift cut in the simulated sample is imposed by the selection of cluster-scale lenses by their lensing efficiency for a $\zs=2$ source plane. On the other hand, the observational SL clusters have been identified using imaging data from various ground- and space-based observatories.
We note that while our simulated sample is statistically inconsistent with the full \citet{Bleem:15} strong lensing sample, considering only lenses at $\zl < 0.7$ a Kolmogorov-Smirnov (KS) test does not reject the hypothesis that the simulated and observed SL samples are drawn from the same underlying distribution (KS-statistic $0.264$, p-value $0.159$).
Regardless, the results presented in this work are not dependent on these samples being drawn from the same underlying distribution.

The redshift range of the simulated SL sample, $z_L \sim 0.16 - 0.67$,  is similar to that of the Sloan Giant Arc Survey (SGAS; M. Gladders et al., in preparation; \citealt{Bayliss:11, Sharon:20}), which identified lensing clusters from giant arcs in shallow optical SDSS imaging. 
Future studies will extend to higher redshifts to complement surveys with samples of galaxy clusters out to $z = 1.75$ such as the SPT-SZ 2500-Square-Degree survey \citep{Bleem:15}.

\subsection{Ray Tracing and Density Maps}
\label{subsec:ray_tracing_sdens}

The ray-traced images and the projected mass distributions of the galaxy clusters have a size of $2048 \times 2048$ pixels and a resolution of $dx = 0\farcs09$ per pixel. For more details of the exact procedure to obtain the lensing maps and the ray-traced images, refer to \citet{Li:16}. Using the surface density distributions of these clusters, we compute all of the lensing maps, including the deflection angle ($\vect{\alpha}$) using Fourier methods, the convergence ($\kappa$), the shear ($\gamma$), the magnification ($\mu$), and the tangential and radial critical curves.

\begin{figure}
\includegraphics[width=0.5\textwidth]{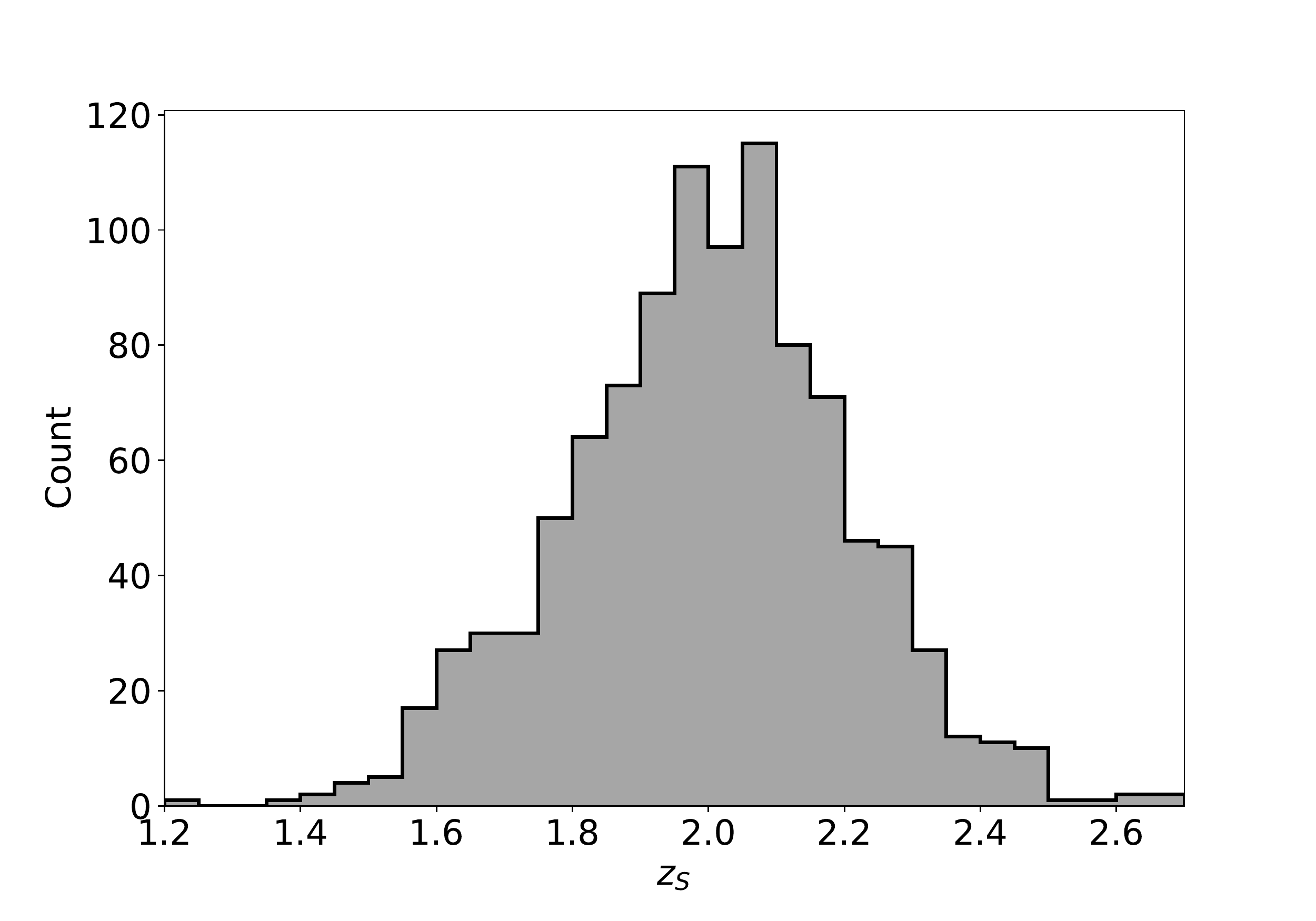}
\caption{\textsc{\textbf{Simulated Background Source Redshifts, $\zs$.}} The distribution is centered at $z=2$, consistent with the observed redshift distribution of highly magnified giant arcs \citep{Bayliss:11}.}
\label{fig:zs_sample}
\end{figure}

We draw redshifts for 1024 background sources from a distribution ranging from $z \sim 1.2 $ to $z \sim 2.7$, following the observed distribution of \citet{Bayliss:11} (shown in \autoref{fig:zs_sample}). The image plane of each cluster was generated multiple times, resulting in $5 - 24$ ray-tracing realizations for each cluster halo. The background sources were randomly placed in areas of high magnification to produce highly magnified (total magnification $> 5$) arcs easily detected from ground based observations (e.g., \citealt{Bayliss:11, Sharon:20}).

We note that the ray-tracing did not take into account structures along the line-of-sight. Structure along the line-of-sight can boost the total number of lenses observed by increasing the SL cross-section of individual clusters, having a larger effect on the less massive primary lensing halos \citep{Puchwein:09, Bayliss:14, Li:19}. The magnification of the arcs is also affected by the structure along the line-of-sight requiring particular care when studying the background source properties \citep{Bayliss:14, DAloisio:14, Chiviri:18} and using strong lensing clusters for cosmological studies \citep{Bayliss:14}. A statistical analysis of how the measurement of the core mass is affected by line of sight structure is left for future work. 

We use the ray-traced images to compute the mass enclosed by the Einstein radius, and the surface density maps as ``true'' mass to characterize the efficacy of this mass estimate.

\section{methodology} 
\label{sec:methods}

Our methodology attempts to mirror the procedures that would be used in SL analyses of real data. Even in large surveys such as SPT, this includes a significant component of manual inspection and identification of SL evidence. Manual inspection is also required for targeted spectroscopic follow-up.

\subsection{Einstein Radius Measurement}
\label{subsec:get_radii}

The first step is to measure an Einstein radius from the positions of the lensed images (arcs). To locate the arcs, we examine each of the ray-traced images by eye to identify sets of multiple images using their morphology and expected lensing geometry, mimicking the process of finding multiply-imaged lensed systems in observational data. If multi-band information is available lens modelers also take advantage of color information of the lensed images, but in this particular case, color information is not available from the ray-traced images.

\begin{figure}
\includegraphics[width=0.5\textwidth]{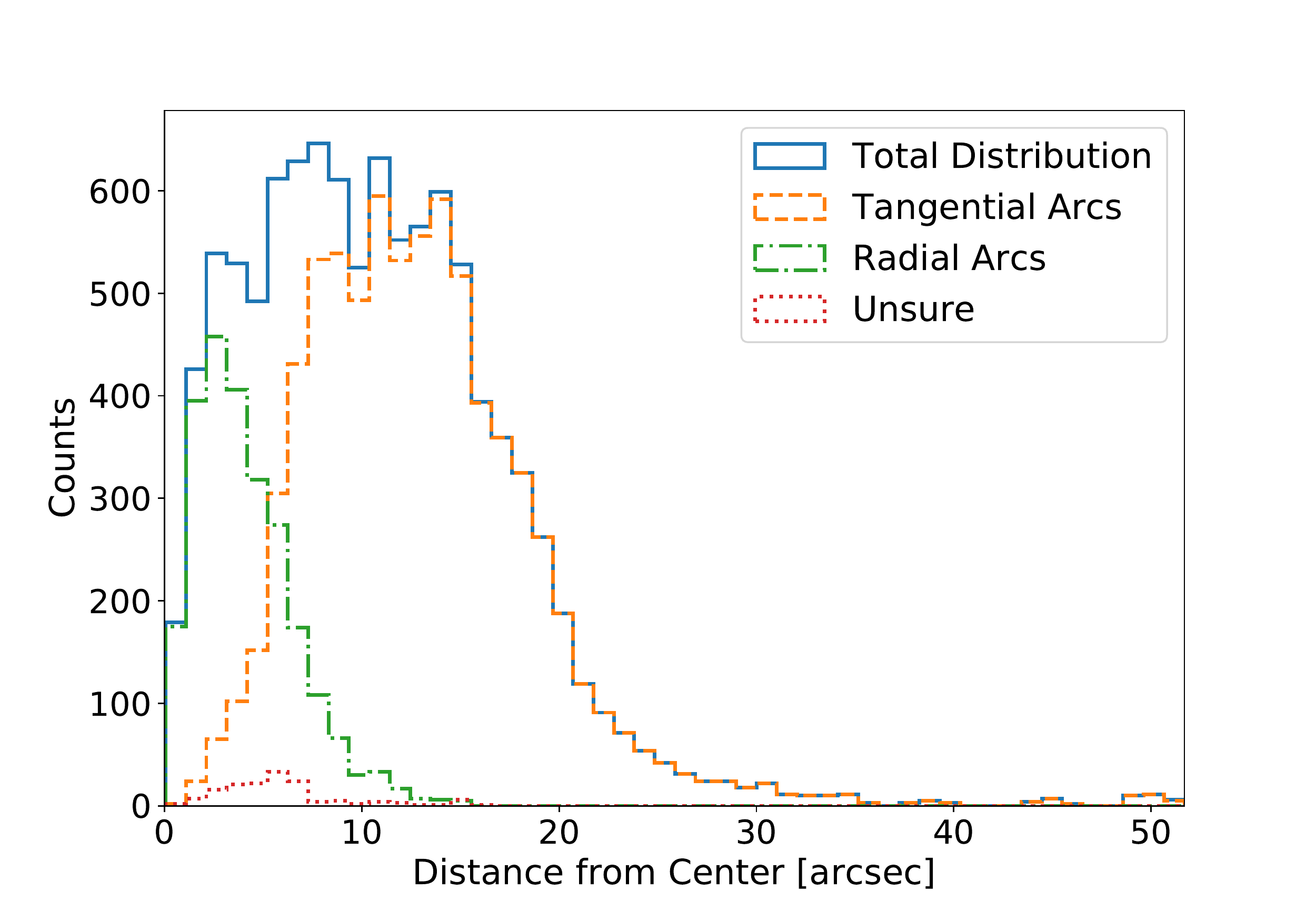}
\caption{\textsc{\textbf{Radial Distribution of the Identified Arcs.}} Radial distances are measured with respect to the pixel with the highest projected mass density of the simulated galaxy cluster. We display the distribution of the tangential arcs with an orange dashed line, radial arcs with a green dashed-dotted line, and those images for which we are unsure with a red dotted line. The distribution of the radial and tangential arcs matches our expectation from lensing geometry, having radial arcs closer to the center while tangential arcs are found farther out.}
\label{fig:arc_dist}
\end{figure}

Using this process, we created a catalog with flags identifying the tangential and radial arcs, corresponding to the tangential and radial critical curves, respectively (see \S \ref{sec:lensing}). Identified lensed images whose classification (radial or tangential) is unclear were noted. The radial distribution of the identified arcs is shown in \autoref{fig:arc_dist}. We find that the distribution of tangential and radial arcs match our expectations from lensing geometry, the radial arcs are found near the center while the tangential arcs are typically found farther out. The distribution we find is qualitatively consistent with \citet{Florian:16}.

Since the Einstein radius is a representation of the tangential critical curve \citep{Bartelmann:10,Kneib:11}, we only include the tangential arcs when finding the Einstein radius. 
We fit a circle to the tangential arcs as explained below; the radii of the resulting circles shall be our Einstein radii, $\theta_E$.

We explore three alternatives for the centering of the circle; in the first method (hereafter \textit{fixed center}) we fix the center of the circle to the point of highest surface density of the projected mass from the simulated halo.
Since in observations we do not a priori know where the center of the dark matter halo is located, in the second method we set the center as a free parameter (hereafter \textit{free center}) with a conservative uniform prior of $\pm 13\farcs5$ from the projected 3-D potential center of the halo. Because the free center requires two more free parameters, the free center minimization was only performed on the cases where 3 or more multiple images were identified as tangential arcs.
In the observational realm, the BCG can be, and often is, used as a proxy for the cluster center. 
The third method (hereafter \textit{fixed center with BCG offset}) mimics fixing the center to an observed BCG.
Since the Outer Rim simulation does not include baryonic information, we cannot determine the BCG position directly from it.  We therefore turn to studies that investigate the BCG offset from the dark matter center. 
\citet{Harvey:19} explores the radial offset between the BCG and the dark matter (DM) center as an observable test of self-interacting dark matter (SIDM) models with different dark matter cross-sections. 
They find that the BCG-DM offset follows a log-normal distribution, with the offsets in the cold dark matter (CDM) case being the smallest ($\mu = 3.8 \pm 0.7$ kpc) and increases with increasing dark matter cross section. We use the distribution of the SIDM model with a DM cross-section of $0.3$ cm$^2$/g. This value represent a reasonable/conservative upper boundary according to recent analysis \citep{Pardo:19, Sagunski:20}. Following this rationale, we fix the center of the circle to a point offset from the center of the dark matter halo, with a radial offset drawn from a log-normal distribution with $\mu = 6.1 \pm 0.7$ kpc, in a random direction.

For the fitting procedure, we use an ensemble sampler Markov chain Monte Carlo (MCMC) implemented for python using the libraries emcee\footnote{Python emcee \url{https://emcee.readthedocs.io/en/stable/}} \citep{Foreman:13} and lmfit\footnote{Python lmfit \url{https://lmfit.github.io/lmfit-py/index.html}} \citep{Newville:14} method to fit a circle to the tangential arcs. The fitting method minimizes the distance between the 2-D position of the arcs (visually identified morphological features that can be matched between the multiple images) and the nearest point to it on the circle. We use a uniform prior in the radius fitting parameter of $2\farcs25 < \theta_E < 45\farcs0$ for all three of our fitting methods. We note that the cases where only a single arc is identified, the distance between the fixed center and the arc is used to determine the radius of the circle and no scatter is measured.

\begin{figure}
\includegraphics[width=0.5\textwidth]{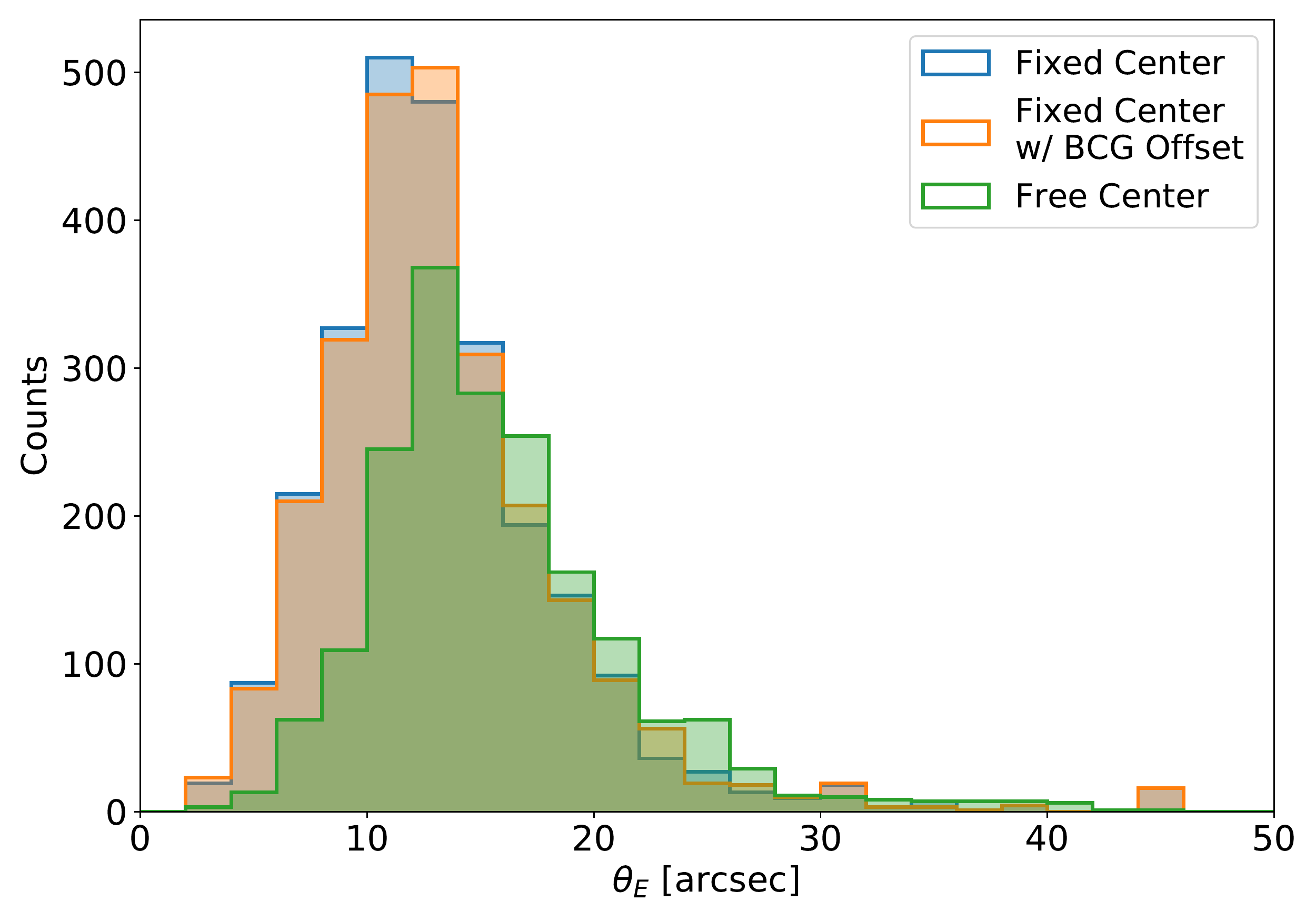}
\caption{Distribution of the Einstein radii from the fits to the identified tangential arcs utilizing the fixed center (blue), fixed center with BCG offset (orange) and free center (green).}
\label{fig:ang_r_dist}
\end{figure}

\begin{figure*}
\center
\includegraphics[width=1\textwidth]{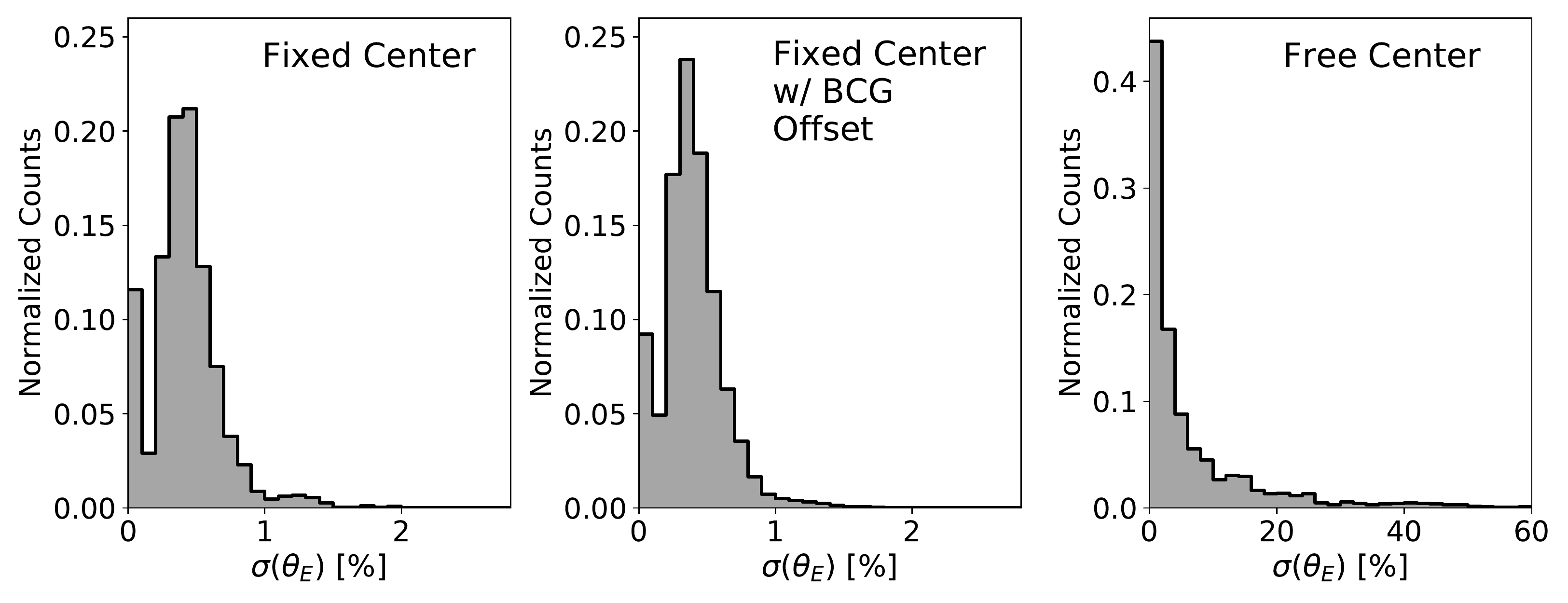}
\caption{Distribution of the standard deviation of the measured Einstein radii ($\sigma (\theta_E)$) in units of percentage utilizing the fixed center (\textit{left}), fixed center with BCG offset (\textit{middle}), and free center  (\textit{right}). We find that the standard deviation of the free center method is about 20 times higher than that of the fixed center and fixed center with BCG offset methods.}
\label{fig:ang_r_unc_dist}
\end{figure*}

The distribution of the measured $\theta_E$ is shown in \autoref{fig:ang_r_dist} and the distribution of the standard deviation, $\sigma (\theta_E)$, computed from the covariance matrix of the fit is shown in \autoref{fig:ang_r_unc_dist}.
Since the free center fitting procedure is significantly more flexible, the standard deviation on the fitted $\theta_E$ for the free center is about 20 times higher compared to that of the fixed center and fix center with BCG offset fit.

\subsection{Inferred Mass}
\label{subsec:get_mass}

\begin{figure*}
\center
\includegraphics[width=1\textwidth]{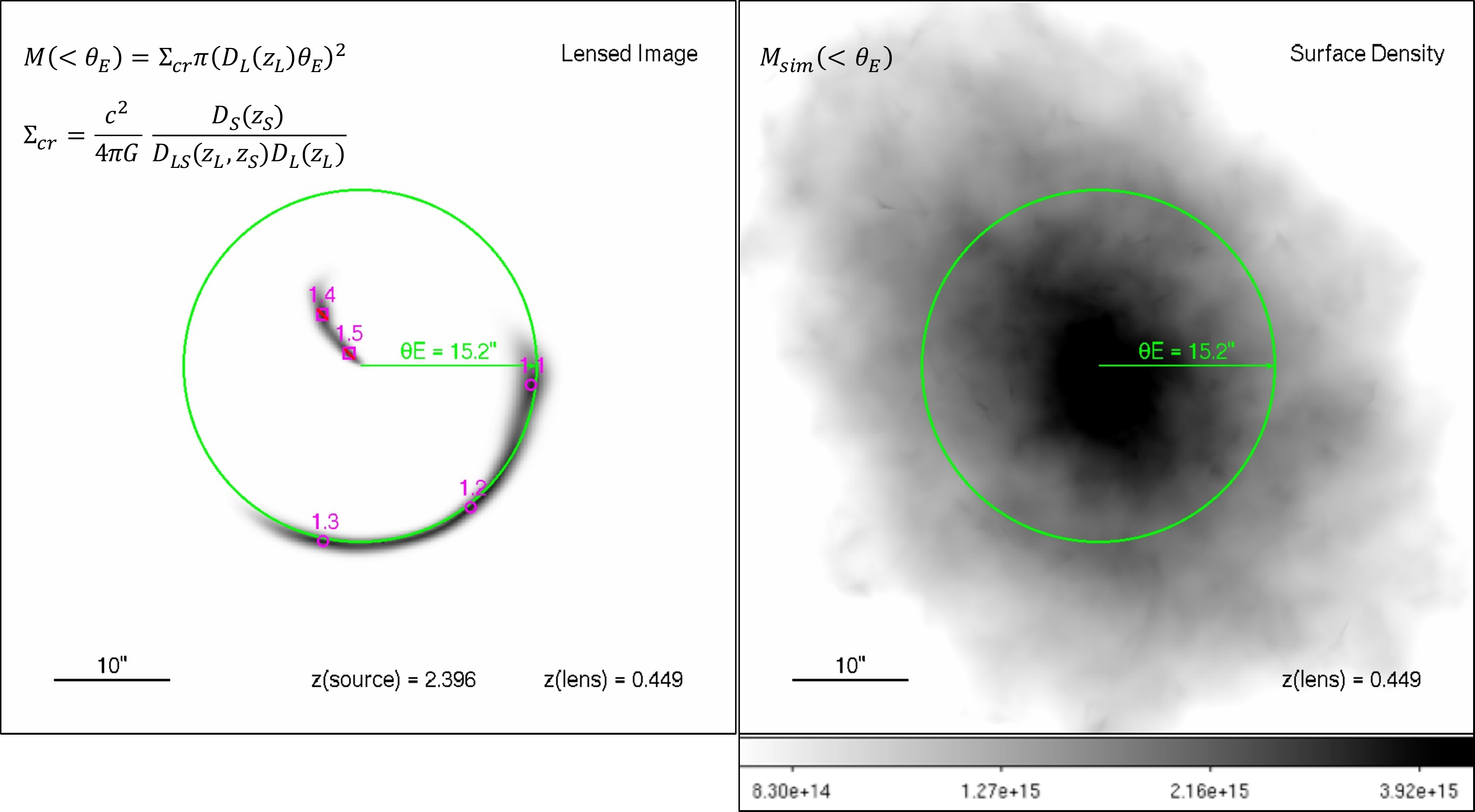}
\caption{\textsc{\textbf{Example of the Simulated Images to Illustrate our Methodology.}} \textit{Left}: ray-traced image; the identified lensed images are indicated with magenta symbols, with circles on tangential arcs and squares with a slash through on radial arcs. We fixed the center to the highest surface density point from the projected mass distribution and fit a circle to the tangential arcs of radius of $\theta_E = 15\farcs2$, shown in green. The mass inferred from the Einstein radius is  $M(<\theta_E) = 3.38 \text{ x } 10^{13} \ \mathrm{M}_\odot \mathrm{h}^{-1}$. \textit{Right}: projected mass density distribution of the simulated galaxy cluster where the green circle is the same aperture from the lensed image. The color-bar is in units of M$_{\odot}$ Mpc$^{-2}$ h. The ``true'' projected mass enclosed is $M_{sim}(<\theta_E) = 3.00 \text{ x } 10^{13} \ \mathrm{M}_\odot \mathrm{h}^{-1}$. We perform our analysis utilizing these two masses, the inferred ($\ML$) and the ``true'' ($\MT$).}
\label{fig:er_explanation}
\end{figure*}

Taking the Einstein radius from \S\ref{subsec:get_radii} and the corresponding lens and source redshifts (\S\ref{subsec:sl_sample}), we compute the enclosed projected mass, $M(<\theta_E)$, via \autoref{eq:m_er}. For our comparison, we use the projected mass distribution from the simulation to measure the true mass enclosed within the same aperture. We refer to this as the ``true'' mass, $M_{sim}(<\theta_E)$. An example of this procedure is shown in \autoref{fig:er_explanation}.

\subsection{Statistical approach to Correctly Represent the Universe}
\label{subsec:stats}

Our simulated sample consists of a total of $1024$ ray tracing realizations through 74 strong lensing galaxy clusters, resulting with $5$-$24$ ray-tracing realizations for each cluster. Each ray-traced simulated realization includes one of the 74 cluster halos and a single background source at a unique redshift. In addition, in some of the realizations multiple distinct structures (clumps) were identified and used to measure more than one Einstein Radius for that particular realization. For this reason the ray-trace realizations and Einstein Radius for a specific galaxy cluster are not independent from each other.  

To establish a robust analysis that represents the universe, includes the statistical uncertainty of the fitted Einstein radius, and allows for the application to observational data, we weight each galaxy cluster to equal one. The ray-traced realizations are then evenly weighted by a factor of one over the total number of realizations for the specific cluster, and similarly the Einstein radii were weighted per ray-traced image. For each galaxy cluster, we select, at random, one ray-traced image from that cluster and one Einstein radius measurement for that realization. We then sample the selected Einstein radius using a normal distribution with the mean as the best fit Einstein radius and standard deviation equal to the uncertainty of the fitted Einstein radius. We repeat this process $1,000$ times per cluster and use this sample with $74,000$ points for our statistical analysis.


\section{Analysis of Results} 
\label{sec:analysis}

In this section, we compare the mass inferred from the Einstein radius ($\ML$) to the true mass ($\MT$), measured from the surface density maps within the same aperture (\autoref{fig:er_explanation}); measure the scatter of this mass estimate; and explore any dependence on the galaxy cluster properties, as well as observational information available from the ray-traced images. 

\begin{figure*}
\center
\includegraphics[width=1\textwidth]{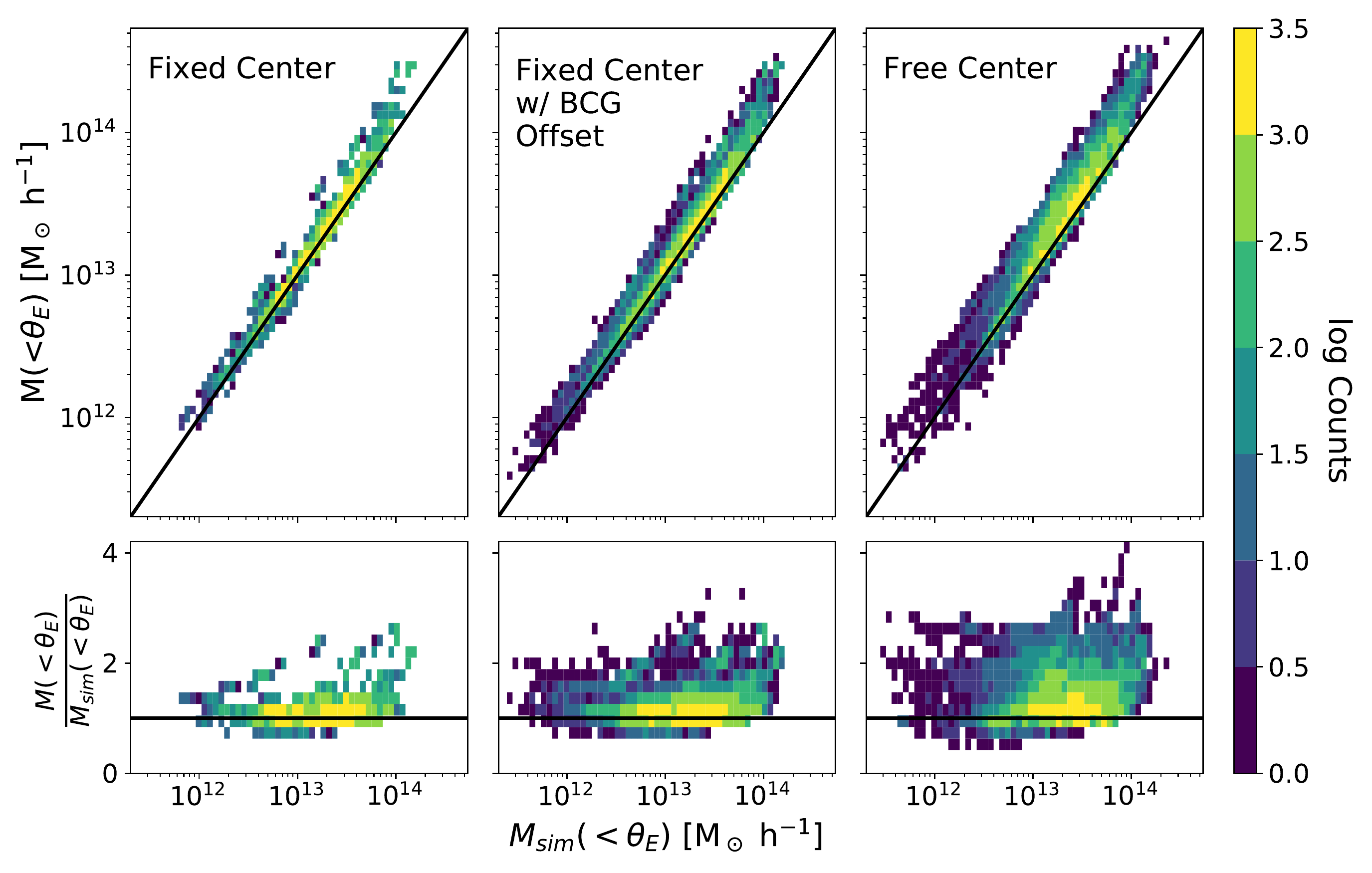}
\caption{\textsc{\textbf{Mass Comparison Between the $\ML$ and $\MT$.}} The mass comparison for the fixed center (\textit{left}), fixed center with BCG offset (\textit{middle}), and free center (\textit{right}) are shown. $\MT$ and $\ML$ are given in units of $\Msun h^{-1}$ and the solid black line is where $\MT = \ML$. The bottom plots show the ratio of the masses, $\ML\ /\ \MT$. The total number of counts is the $74,000$ sampled data points (\S\ref{subsec:stats}) used in the analysis of the scatter and bias of the $\ML$ compared to $\MT$. We find that $\ML$ overestimates $\MT$ in all cases, especially at large masses, and the scatter is smallest for the fixed center method and highest for the free center method.}
\label{fig:MvM}
\end{figure*}

In \autoref{fig:MvM}, we show a direct comparison between $\ML$ and $\MT$ for the fixed center (left panel), fixed center with BCG offset (middle panel), and free center (right panel) cases. We find that $\ML$ overestimates $\MT$ in all cases, especially at large masses.

We measure an overall scatter of \FXAllScatter\ and bias of \FXAllBias\ for the fixed center, scatter of \FXAllBCGScatter\ and bias of \FXAllBCGBias\ for the fixed center with BCG offset, and scatter of \FRAllScatter\ and bias of \FRAllBias\ for the free center. The scatter is defined as half the difference between the 84th percentile and the 16th percentile of the distribution and the bias is determined using the median of the distribution.
We note that previous estimates of the uncertainty associated with this measurements state $\sim 30 \%$ \citep{Bartelmann:96, Schneider:06b}, however, it is unclear how the uncertainty is defined.

Comparing the results of the three methods, we find that the free center method is the least reliable in recovering the true mass. Its measured $\theta_E$ statistical uncertainty is 20 times higher than those of the fixed center (\autoref{fig:ang_r_unc_dist}), and the scatter and bias in $\ML\ /\ \MT$ are significantly higher (\autoref{fig:MvM}).
In addition, the free center method is limited to cases where $3$ or more tangential arcs are identified. For these reasons, we do not recommend that the free center method be utilized to measure the Einstein radius and the mass enclosed by the Einstein radius. The fixed center with BCG offset shows that the additional scatter due to the offset between the BCG and dark matter center is small, justifying the use of the observed BCG as the fixed center of the Einstein radius.
For the rest of the paper we are only going to consider the fixed center and the fixed center with BCG offset.

\begin{figure}
\includegraphics[width=0.5\textwidth]{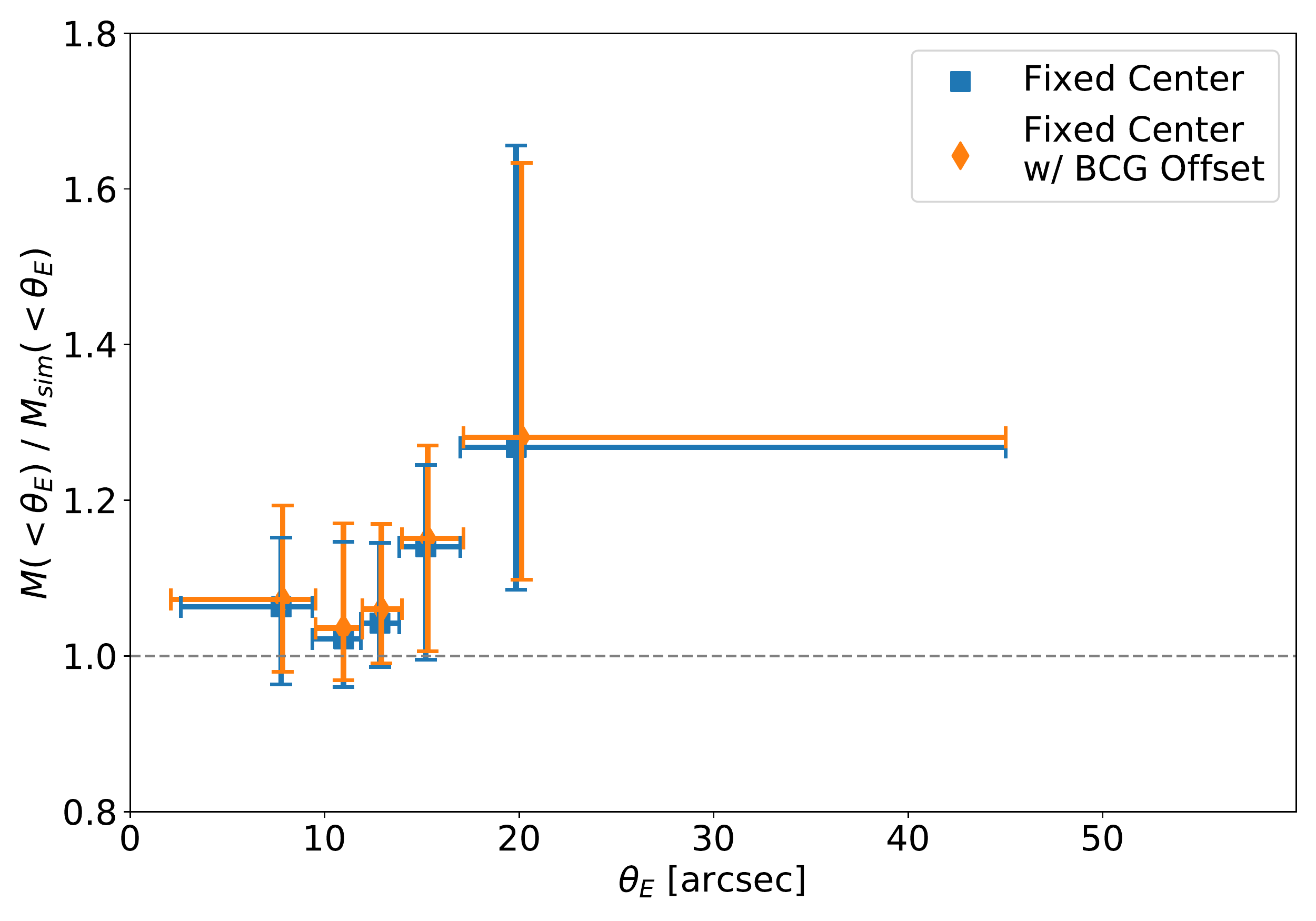}
\caption{\textsc{\textbf{Ratio of inferred to ``true'' mass, $\ML / \MT$, with respect to $\theta_E$}}. The fixed center (blue square) and fixed center with BCG offset center (orange diamond), are shown. The symbol marks the median of the distribution of the mass ratio, the horizontal error bars indicate the bin size, and the vertical error bars represent the $16$th and $84$th percentile. We find a positive bias in all of the bins and that both fixed center and fixed center with BCG offset yield a similar $\theta_E$.}
\label{fig:er_binned_analysis}
\end{figure}

To explore the dependence of this mass estimate on lens properties, we consider the ratio of inferred to true mass,  $\ML / \MT$, and group the measurements into bins of equal number of points. We plot $\ML/ \MT$ with respect to the Einstein radius in \autoref{fig:er_binned_analysis}. This figure shows clearly that  the $\ML$ mass estimate is not randomly scattered about the true mass, and that it overestimates the true mass at all radii. In \S \ref{sec:emp_cor}, we describe an empirical correction to de-bias the measurement of the mass enclosed by the Einstein radius.

In the following sections, we explore possible causes, and identify observable indicators of the scatter and bias of $\ML$.

\subsection{Possible causes and indicators of the scatter in the $\ML$ mass estimate}
\label{subsec:analysis_systematics}

We explore possible dependence of the scatter and bias on $\ML$ with respect to galaxy cluster properties, background source, and lensing geometry. The galaxy cluster properties used in our analysis include: galaxy cluster redshift ($\zl$), total mass ($M_{200}$), concentration ($c_{200}$), dynamical state, and the shape of the tangential critical curve. The total mass, concentration, and dynamical state information for the simulated cluster sample are adopted from \citet{Child:18}. From the background source, we have the redshift information ($\zs$) and from the lensing geometry, we measure how much of the Einstein circle is covered by the arcs ($\phi$), as we explain below.

\paragraph{Lens and Source Redshifts} The redshifts of the lens and the source determine the lensing geometry of the system through the angular diameter distances (\autoref{eq:lenseq}). Redshifts can be determined observationlly, when spectroscopic or extensive photometric information is available. The redshift distribution of the simulated clusters ($\zl$) from the Outer Rim  and background source redshift ($\zs$) are shown in \autoref{fig:cluster_sample} and \autoref{fig:zs_sample}, respectively.

\paragraph{Total Mass and Concentration} $M_{200}$ and $c_{200}$ are adopted from \citet{Child:18}. The distribution of the simulated galaxy cluster total mass and concentration are shown in the left panels of \autoref{fig:cluster_sample}. We note that $M_{200}$ and $c_{200}$ are not directly available from the imaging data at the core of the cluster where the strong lensing evidence is present. However, since our aim is to use the core mass to inform the mass-concentration relation, it is important to test whether this mass estimator introduces correlated bias. 

\begin{figure*}
\center
\includegraphics[width=1\textwidth]{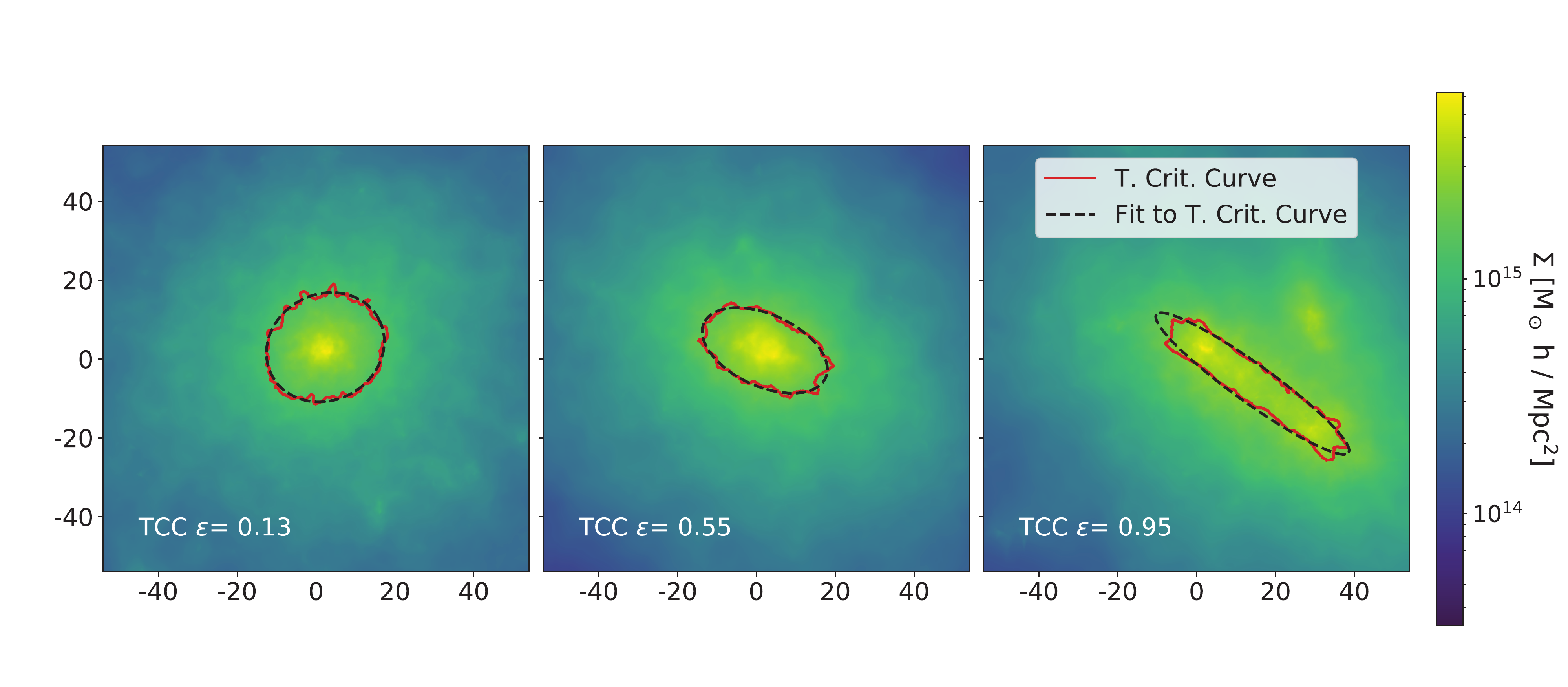}
\caption{\textsc{\textbf{Examples of the ellipticity ($\epsilon$) of the tangential critical curve (TCC) as a proxy for the cluster deviation from spherical symmetry.}} We show as example three simulated clusters with different projected ellipticities. The red line is the tangential critical curve for a particular background source redshift $\zs$. The dashed black line indicates the ellipse fitted to the tangential critical line, from which we compute the ellipticity, $\epsilon$. The lines are plotted over the projected mass distribution of the corresponding simulated galaxy clusters. The x- and y- axes are in units of arcseconds. The color bar indicates the surface density value in units of $\mathrm{M}_\odot \mathrm{h} / \mathrm{Mpc}^{2}$.}
\label{fig:ell_example}
\end{figure*}

\begin{figure}
\includegraphics[width=0.5\textwidth]{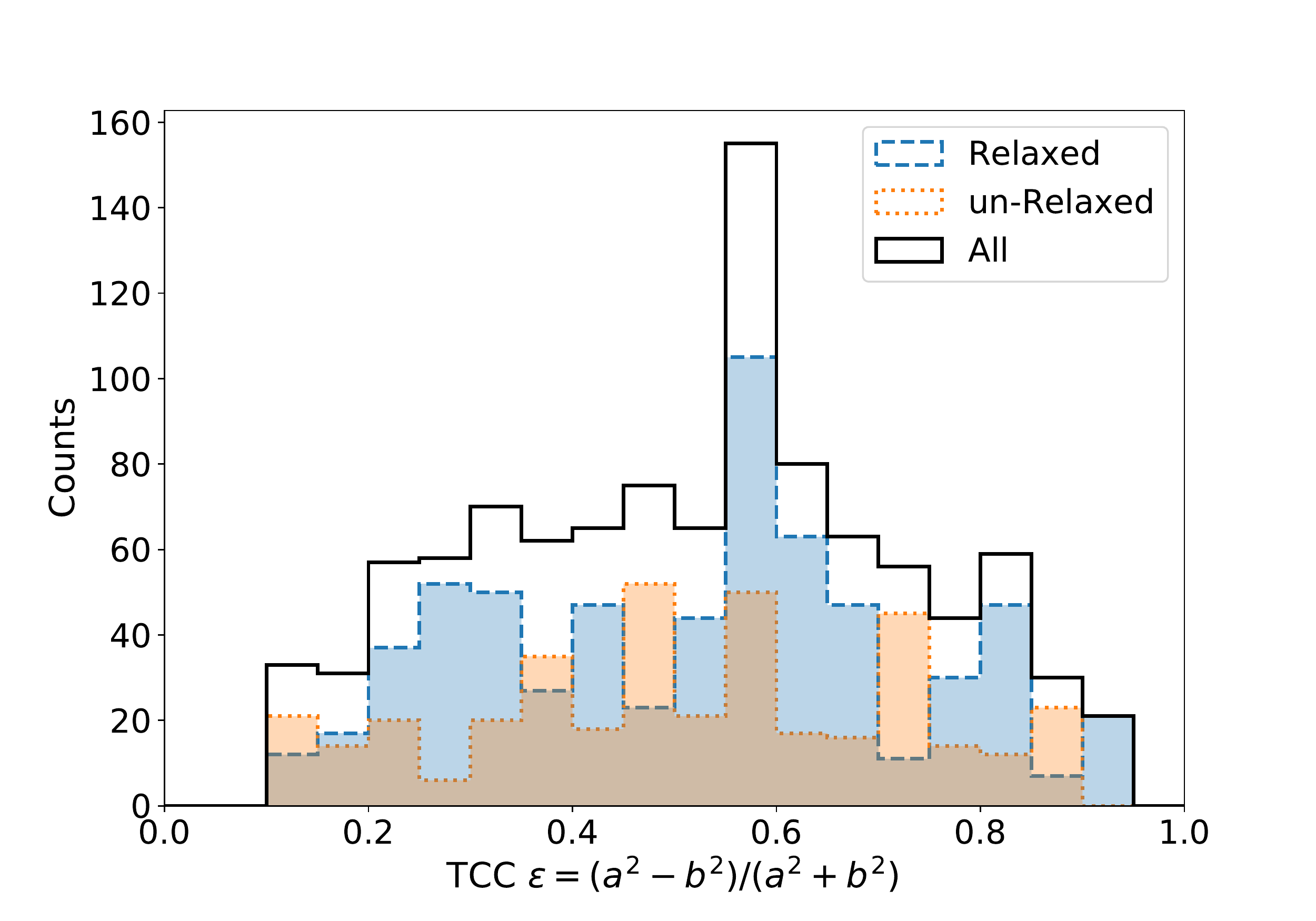}
\caption{\textsc{\textbf{Dynamical State and Deviation from Circular Symmetry.}} Distribution of the tangential critical curve (TCC) ellipticity, $\epsilon$. The overall distribution is indicated by the black line and the contributions from the dynamical (relaxed or un-relaxed) state of the simulated galaxy clusters (from \citet{Child:18}) is indicated by the shaded bars. We observe that the dynamical state information is not an indicator of deviations from spherical symmetry of the simulated galaxy cluster.}
\label{fig:ell_rel}
\end{figure}

\paragraph{Cluster Deviation from Spherical Symmetry} Since galaxy clusters do not have a circular projected mass distribution, we expect differences between $\MT$ and $\ML$ due to deviations from the assumed circular symmetry. To assess the deviation of the lens from spherical symmetry, we use the tangential critical curves derived from the simulation as a proxy for the shape of the projected mass distribution at the core of the cluster. We sample the tangential critical curves with a few hundred to thousands of points by using the python library matplotlib.contour \footnote{Python matplotlib.contour \url{https://matplotlib.org/3.1.0/api/contour_api.html}} setting a contour level at $0$ for the inverse magnification due to the tangential critical curve. Using the technique described in \citet{Fitzgibbon:96}, we fit an ellipse to each tangential critical curve corresponding to every background source redshift. We then use the resultant ellipticity, defined as $\epsilon = (a^2-b^2)/(a^2+b^2)$, where $a$ is the semi-major axis and $b$ is the semi-minor axis. In \autoref{fig:ell_example}, we show three examples of the ellipse fits to the tangential critical curve, over-plotted on the projected mass density distribution. We plot the distribution of ellipticity of the tangential critical curve in \autoref{fig:ell_rel}. This characterization of the projected shape of the galaxy cluster is not accessible directly from the observational data prior to a detailed SL model which this method aims to avoid.

\paragraph{Galaxy Cluster Relaxation State} We tested whether the relaxation state of the galaxy clusters (see \S \ref{subsec:sl_sample} for the simulated sample dynamical state description) can be used as a proxy for the deviation from spherical symmetry. Observationally, this can be determined from X-ray imaging (e.g., \citealt{Mantz:15}). In \autoref{fig:ell_rel}, we plot $\epsilon$ separated by the relaxation state of the galaxy cluster. We perform a two sample Kolmogorov-Smirnov test to quantify the difference between the two samples with a confidence level of $99.7\%$. The KS-statistic is \KSStat\ with a p-value of \KSPvalue. With this test, we cannot reject the null hypothesis that the two samples are drawn from the same continuous distribution. From our KS test and \autoref{fig:ell_rel}, we find no correlation between the dynamical state and $\epsilon$. 

\begin{figure*}
\center
\includegraphics[width=1\textwidth]{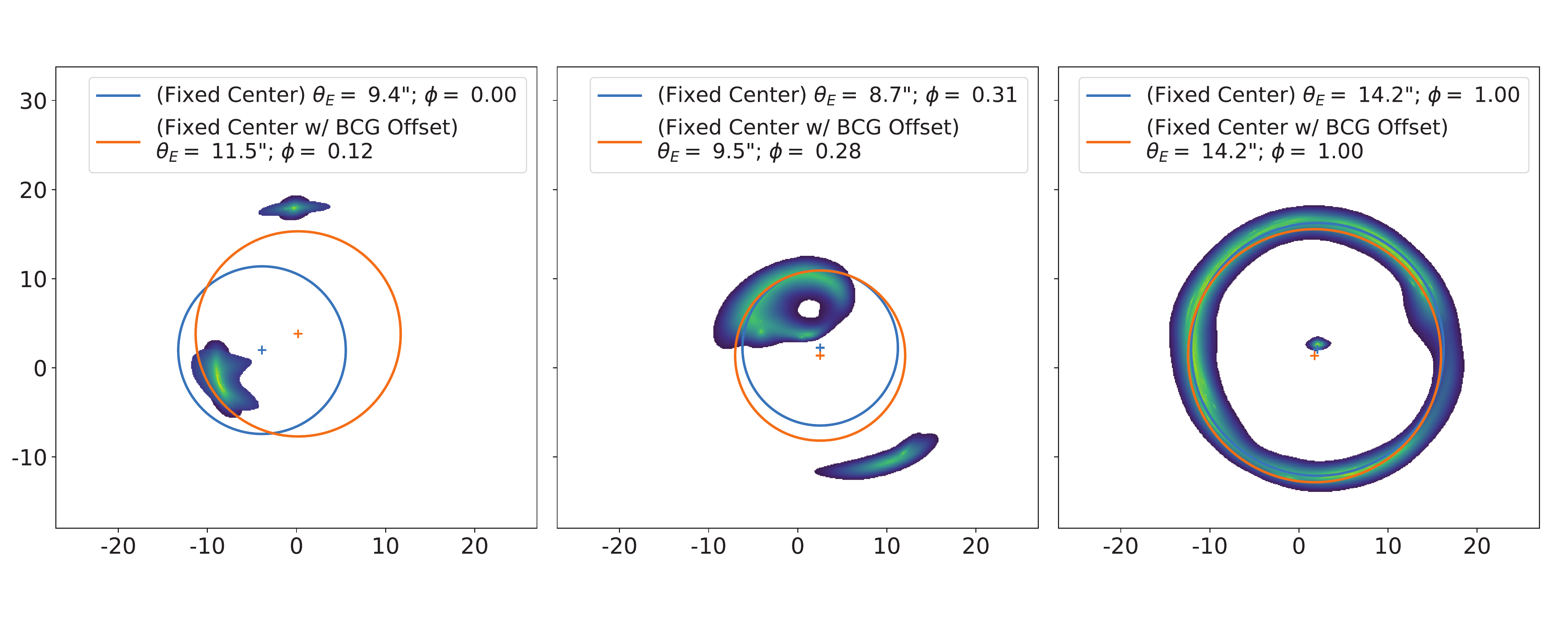}
\caption{\textsc{\textbf{The fraction of circle covered by the arcs ($\phi$) for three examples cases.}} The Einstein radius fitted to the identified tangential arcs for both the fixed center (blue) and one example of the fixed center with BCG offset (orange) are plotted; the corresponding centers of the circles are indicated by the crosses. The BCG offset was determined by drawing a radial offset between the BCG and dark matter halo from a log-normal distribution with $\mu = 6.1 \pm 0.7$ kpc \citep{Harvey:19} and an angle from a uniform distribution form $0$ to $359$ degrees.  The fraction of the circle covered by the arcs for the fixed center and fixed center with BCG offset is shown in the legend. The x- and y-axis are in units of arcseconds.}
\label{fig:frac_arc_example}
\end{figure*}

\begin{figure}
\includegraphics[width=0.5\textwidth]{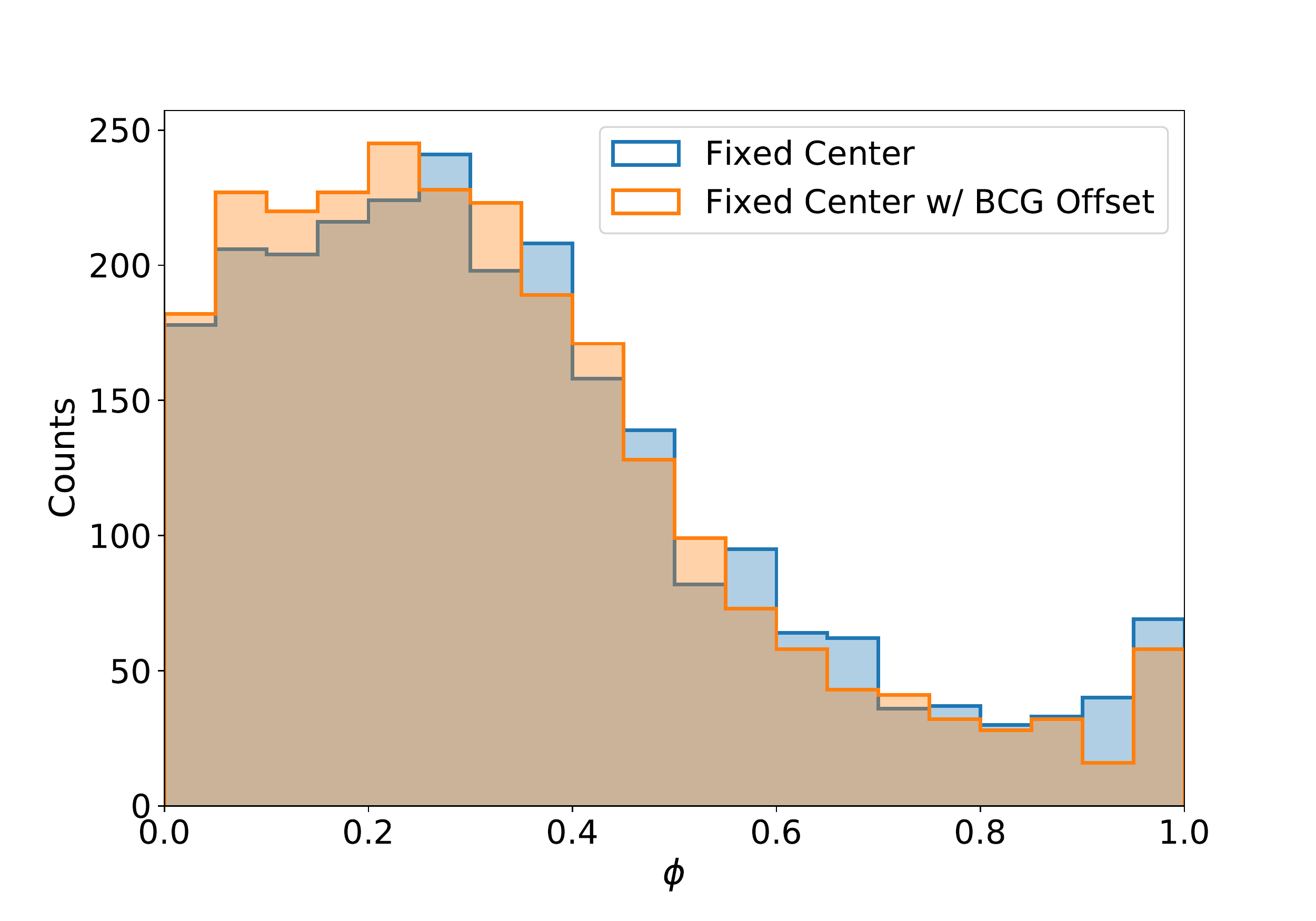}
\caption{\textsc{\textbf{Distribution of the fraction of the circle covered by arcs ($\phi$) for a given source.}} }
\label{fig:frac_arc_dist}
\end{figure}

\paragraph{The fraction of the Einstein circle covered by arcs of an individual lensed source}    \ $\phi$ represents the fraction of the Einstein circle that is covered by arcs of a given source. This property is easily accessible from the imaging data. In \autoref{fig:frac_arc_example}, we show three examples of lensed images plotted with their corresponding Einstein circles fitted using the identified tangential arcs for both the fixed center (blue) and an example of one of the realizations of a fixed center with BCG offset (orange). We plot in \autoref{fig:frac_arc_dist} the distribution of $\phi$ for both the fixed center (blue) and fixed center with BCG offset (orange).

\subsection{Results of the Analysis of Systematics}
\label{subsec:systematic_results}

\begin{figure*}
\center
\includegraphics[width=1\textwidth]{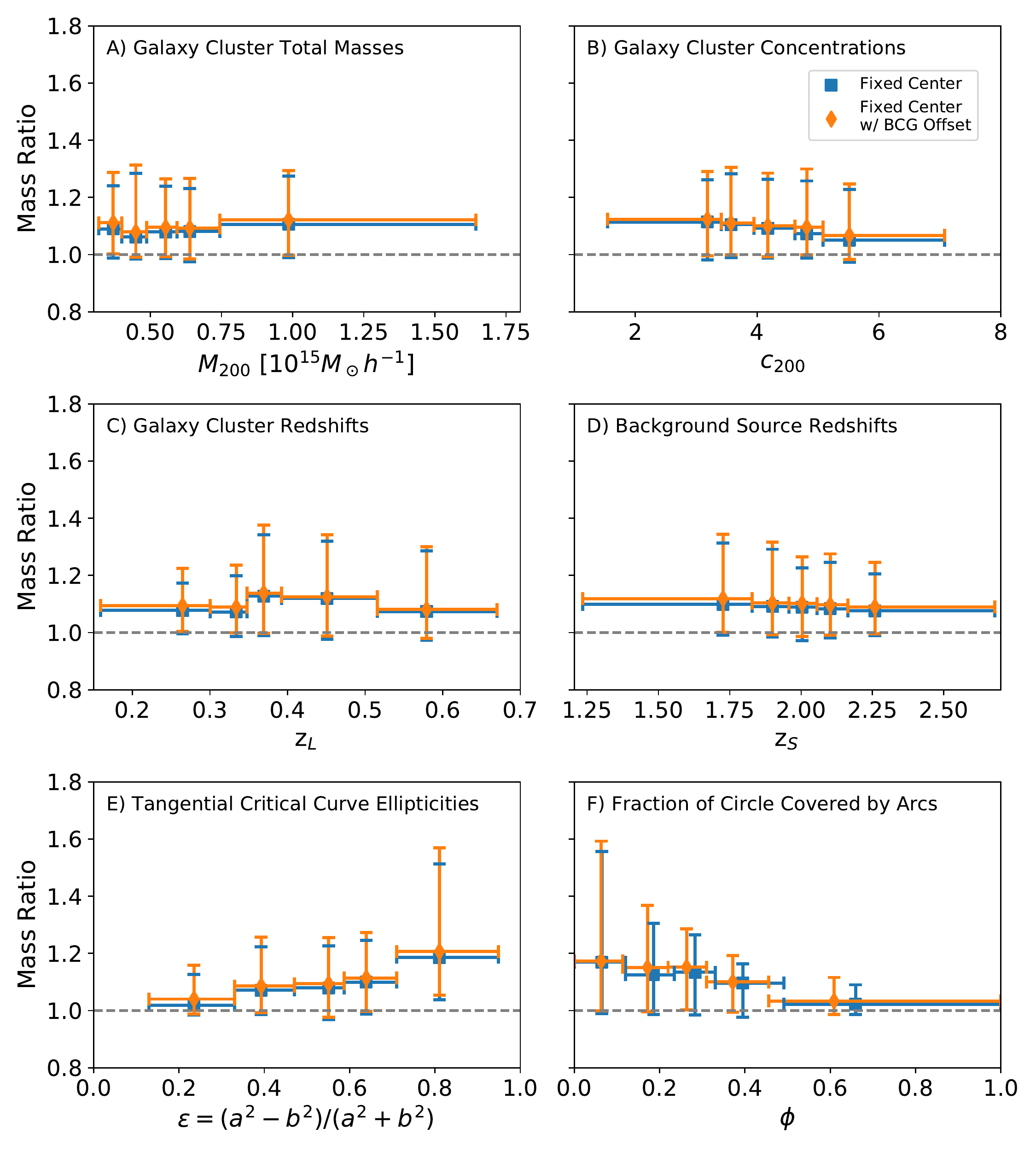}
\caption{\textsc{\textbf{Ratio of Inferred to ``True'' Mass ($\ML / \MT$) Binned by Galaxy Cluster Properties, Background Source, and Lensing Geometry.}} Mass ratio binned by total mass ($M_{200}$, panel A), concentration ($c_{200}$, panel B), cluster redshift ($\zl$, panel C), background source redshift ($\zs$, panel D), tangential critical curve ellipticity ($\epsilon$, panel E), and fraction of circle covered by arcs ($\phi$, panel F). We show results for both the fixed center (blue square) and the fixed center with a BCG offset (orange diamond). The symbol marks the median of the distribution, the horizontal and vertical error bars indicate the bin size and scatter (the 16th and 84th percentile of the distribution), respectively. We find that there is a positive bias in all of the bins. We observe a clear trend with $\epsilon$, where both the scatter and bias increase with increasing $\epsilon$, and $\phi$, where both the scatter and bias decrease as $\phi$ increases.}
\label{fig:binned_analysis}
\end{figure*}

We split the measurements of $\ML$ into equal bins of $M_{200}$, $c_{200}$, $\epsilon$, $\zl$, $\zs$, and $\phi$ and check whether the bias and scatter in the $\ML$ mass estimate depend on these properties. We find that the scatter and bias of $\ML/\MT$ do not depend on four of these properties: the total mass, concentration, cluster redshift, and source redshift, showing flat and uniform progression in panels A--D of \autoref{fig:binned_analysis}. We also note, we find no difference in the bias and scatter of $\ML/\MT$ between the relaxed and un-relaxed clusters nor a correlation between the relaxation state and the bias and scatter of $\ML/\MT$.

Conversely, there are strong correlations between the scatter and bias with respect to the ellipticity of the tangential critical curve ($\epsilon$) and the fraction of the circle covered by arcs ($\phi$).
As can be seen in panel E of \autoref{fig:binned_analysis}, as $\epsilon$ increases both the scatter and bias increase. The dependence on the ellipticity is expected, since one of the main assumptions in the $\ML$ formalism is circular symmetry ($\epsilon = 0.0$). Unfortunately, the measurement of the ellipticity of the tangential critical curve cannot be determined until after a lens model has been computed.

The scatter and bias of $\ML$ decrease with increasing $\phi$ (\autoref{fig:binned_analysis}, panel F). This trend matches our expectation; lenses with $\phi$ closer to $1.0$ are typically more circular. Unlike the ellipticity, the fraction of the fitted circle covered by arcs is readily available from the same data used for analysis of observed clusters. It is therefore a useful estimator of lens-dependent uncertainty. For convenience, we tabulate the information displayed in Panel F of \autoref{fig:binned_analysis}, in \autoref{table:frac_arc_binned} in the Appendix.


\section{The Effect of Background Source Redshift}
\label{sec:no_zs}

The redshifts are a piece of information that ideally would be available to the lensing analysis, coming from spectroscopic follow-up (e.g., \citealt{Sharon:20}) or using photometric redshifts (e.g., \citealt{Molino:17, Cerny:18}) from extensive multi-band photometry. However, this may not always be the case, especially considering future large surveys where follow-up may be incomplete. We therefore investigate the additional scatter in the mass estimate due to an unknown source redshift. In this analysis, we assume that we know the underlying distribution of the background source redshifts \citep{Bayliss:11}.

To evaluate this case, we use the Einstein radius from \S\ref{subsec:get_radii} and the lens redshift from \S\ref{subsec:sl_sample}, but instead of using the actual source redshifts, we draw $10,000$ source redshifts from a normal distribution with $\mu = 2.00$ and $\sigma = 0.2$. 

\begin{figure*}
\center
\includegraphics[width=1\textwidth]{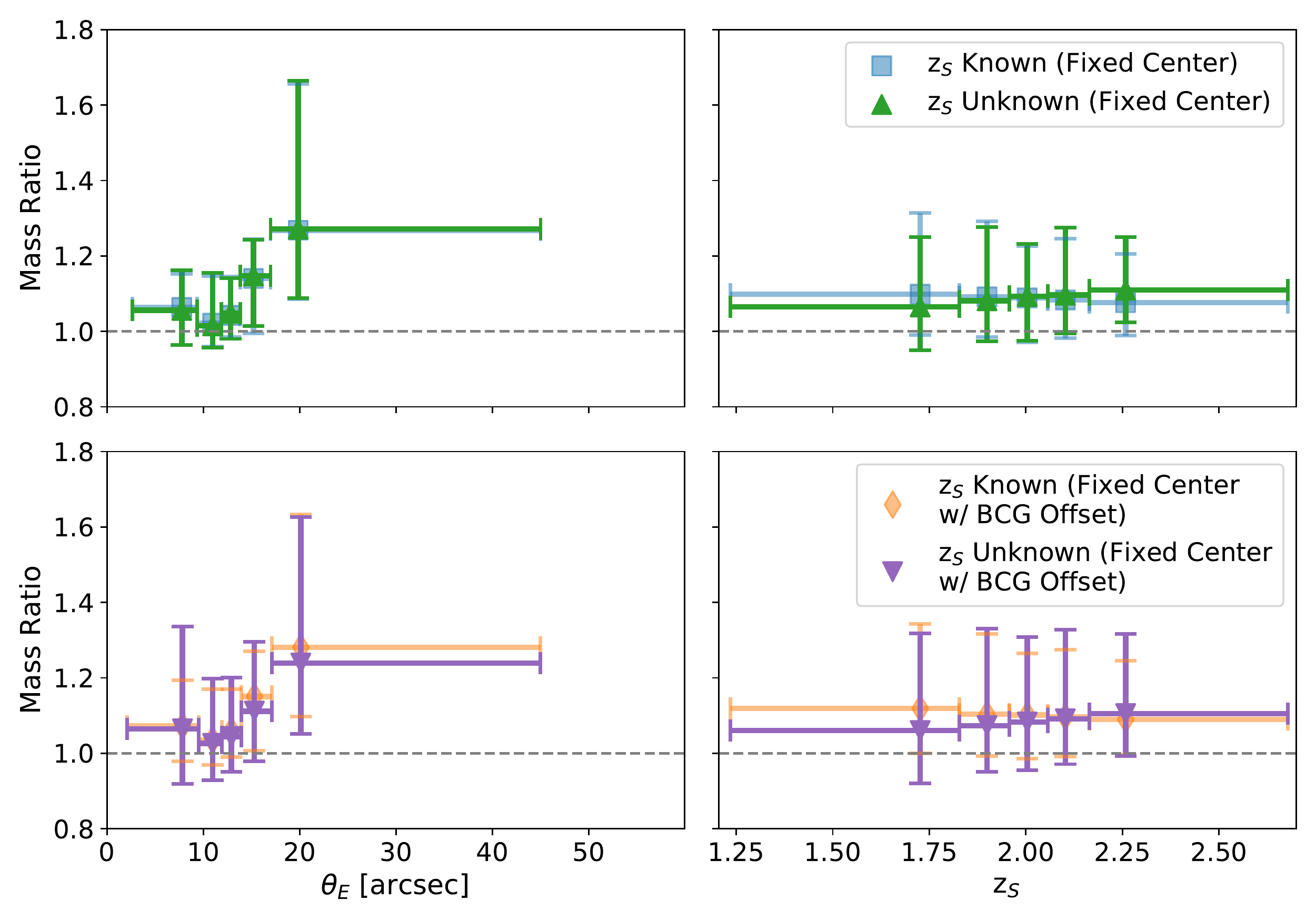}
\caption{\textsc{\textbf{The Effect of Source Redshift Uncertainty on the Results.}} The blue square symbols and orange diamonds represent the fixed center and fixed center with BCG offset, and are the same as \autoref{fig:er_binned_analysis} and \autoref{fig:binned_analysis}, Panel D, respectively. The ratio of the inferred to ``true'' mass for the unknown source redshift are indicated with up-pointing, green, triangles and down-pointing, magenta, triangles. We find that the uncertainty in source redshift has small effects on the results. As expected, when binned by source redshift (\textit{right}), we find that the inferred mass is low at $\zs <2.0$ and high at $\zs>2.0$.}
\label{fig:nozs_binned_analysis}
\end{figure*}

We repeat the analysis in \S \ref{sec:analysis} with this set of drawn background source redshifts. In \autoref{fig:nozs_binned_analysis}, we plot the ratio of the inferred to ``true'' mass in bins of Einstein radius (left panels) and true background source redshift (right panels). We plot the results for both the fixed center (top panels) and the fixed center with BCG offset (bottom panels). For comparison, we over-plot the results from \S~\ref{subsec:systematic_results}. We compute a scatter of \FXNoZScatter\ (\FXNoZBCGScatter) and bias of \FXNoZbias\ (\FXNoZBCGbias) for the fixed center (fixed center with BCG offset).

As can be seen in the left panel of \autoref{fig:nozs_binned_analysis} and the scatter and bias of the fixed center, not knowing the exact background redshift and assuming a normal distribution with $\mu = 2.00$ and $\sigma = 0.2$ for typical giant arcs introduces a negligible uncertainty, particularly when compared to the magnitude of the systematics presented in \S \ref{sec:analysis}. 
Split by bins of background source redshift, the scatter remains the same, however, the inferred mass is higher if $\zs>2$ and lower if $\zs <2$. 

It is important to note that precise source redshifts are critical for most applications of strong lensing (e.g., magnification, time delay, and detailed mass maps). They become negligible in this case because the total enclosed mass is a particularly robust measurement, and the goal is determining the mass of a statistical sample. For mass estimates of individual systems, since the dependence on redshift is straight forward (see \autoref{eq:m_er}) the uncertainties can be easily determined. 


\section{Empirical Corrections}
\label{sec:emp_cor}

As can be seen in Figures \ref{fig:er_binned_analysis} and \ref{fig:nozs_binned_analysis}, the scatter and bias of this estimator shows dependence on $\theta_E$. We explore the use of an empirical correction to un-bias the mass estimate and reduce the scatter obtained from the Einstein radius method. 

We bin the $74,000$ data points into $25$ bins with equal number of data points per bin, using the Doane's formula \citep{Doane:76} to determine the number of bins for a non-normal distribution. We fit a linear, quadratic, and cubic models to the median of the mass ratio ($\ML / \MT$) in each bin and the center of the bin, using the Levenberg-Marquardt algorithm \citep{Levenberg:44, Marquardt:63}. 
We compute the Bayesian Information Criterion (BIC) for each model \citep{Schwarz:78, Liddle:07}. The results of the fits can be found in \autoref{table:emp_corr_models} including the scatter and bias of the resulting empirically corrected data. The BIC results for the fixed center (fixed center with BCG offset) are $-125.7\ (-126.5)$ for the linear, $-152.1\ (-157.2)$ for the quadratic, and $-150.7\ (-156.9)$ for the cubic model. Based on this criterion, the quadratic fit, which has the lowest BIC, is clearly preferred over linear and slightly over cubic fits. We therefore use the quadratic fit to determine an empirical correction:

\begin{equation}
    \frac{\ML}{\MT} = \mathrm{B} \theta_E^2 + \mathrm{C} \theta_E + \mathrm{D} \equiv {f(\theta_E)},
    \label{eq:quad_eq}
\end{equation}

\noindent where B, C, and D are the fit parameters.

\capstartfalse
\begin{deluxetable*}{lccccccc}
\tablecolumns{8}
\tablewidth{2\columnwidth}
\tablecaption{Empirical Correction Models.}
\tablehead{Model & A[arcsec$^{-3}$] & B[arcsec$^{-2}$] & C[arcsec$^{-1}$] & D & BIC & Scatter & Bias}
\startdata
Fixed Center\\
\hline
Cubic & $-4.34\times10^{-5} \pm 3.36\times10^{-5}$ & $3.71\times10^{-3} \pm 1.73\times10^{-3}$ & $-0.06 \pm 0.03$ & $1.29 \pm 0.13$ & $-150.7$ & $10.0$\% & $-0.2$\%\\
Quadratic & --- & $1.49\times10^{-3} \pm 2.11\times10^{-4}$ & $-0.02 \pm 7.05\times10^{-3}$ & $1.14 \pm 0.05$ & $-152.1$ & $10.1$\% & $-0.4$\%\\ 
Linear & --- & --- & ~~$0.02 \pm 2.92\times10^{-3}$ & $0.79 \pm 0.04$ & $-125.7$ & $11.4$\% & $-0.5$\%\\
\hline 
\multicolumn{2}{l}{Fixed Center w/ BCG Offset}\\
\hline
Cubic & $-4.52\times10^{-5} \pm 2.84\times10^{-5}$ & $3.81\times10^{-3} \pm 1.48\times10^{-3}$ & $-0.06 \pm 0.02$ & $1.31 \pm 0.11$ & $-156.9$ & $10.8$\% & $-0.2$\%\\
Quadratic & --- & $1.47\times10^{-3} \pm 1.84\times10^{-4}$ & $-0.02 \pm 6.25\times10^{-3}$ & $1.15 \pm 0.05$ & $-157.2$ & $10.9$\% & $-0.3$\%\\ 
Linear & --- & --- & ~~$0.02 \pm 2.84\times10^{-3}$ & $0.81 \pm 0.04$ & $-126.5$ & $12.1$\% & $-0.4$\%\\
\enddata
\tablenotetext{}{Model fit results of an empirical correction to un-bias and decrease the scatter of the mass enclosed by the Einstein radius. The last two columns are the scatter and bias of the empirically corrected data. The ``fixed center with BCG offset'' analysis accounts for the uncertainty added by using the BCG as a proxy for cluster center.}
\label{table:emp_corr_models}
\end{deluxetable*}
\capstarttrue

We choose not to include $\phi$ in our empirical correction because the parameter is dependent on the resolution of the telescope, depth of the observations, and observing conditions. The value of $\phi$ varies from observation to observation and therefore having a coarser estimate using the binned value in \autoref{table:frac_arc_binned} is more appropriate. We correct the measured $\ML$ by dividing it by the corresponding value computed from the parabolic equation evaluated at $\theta_E$:

\begin{equation}
    \mathrm{Corrected}\ \ML = \mathrm{Measured}\ \ML / f(\theta_E).
    \label{eq:un_bias_ml}
\end{equation} 

\begin{figure}
\includegraphics[width=0.5\textwidth]{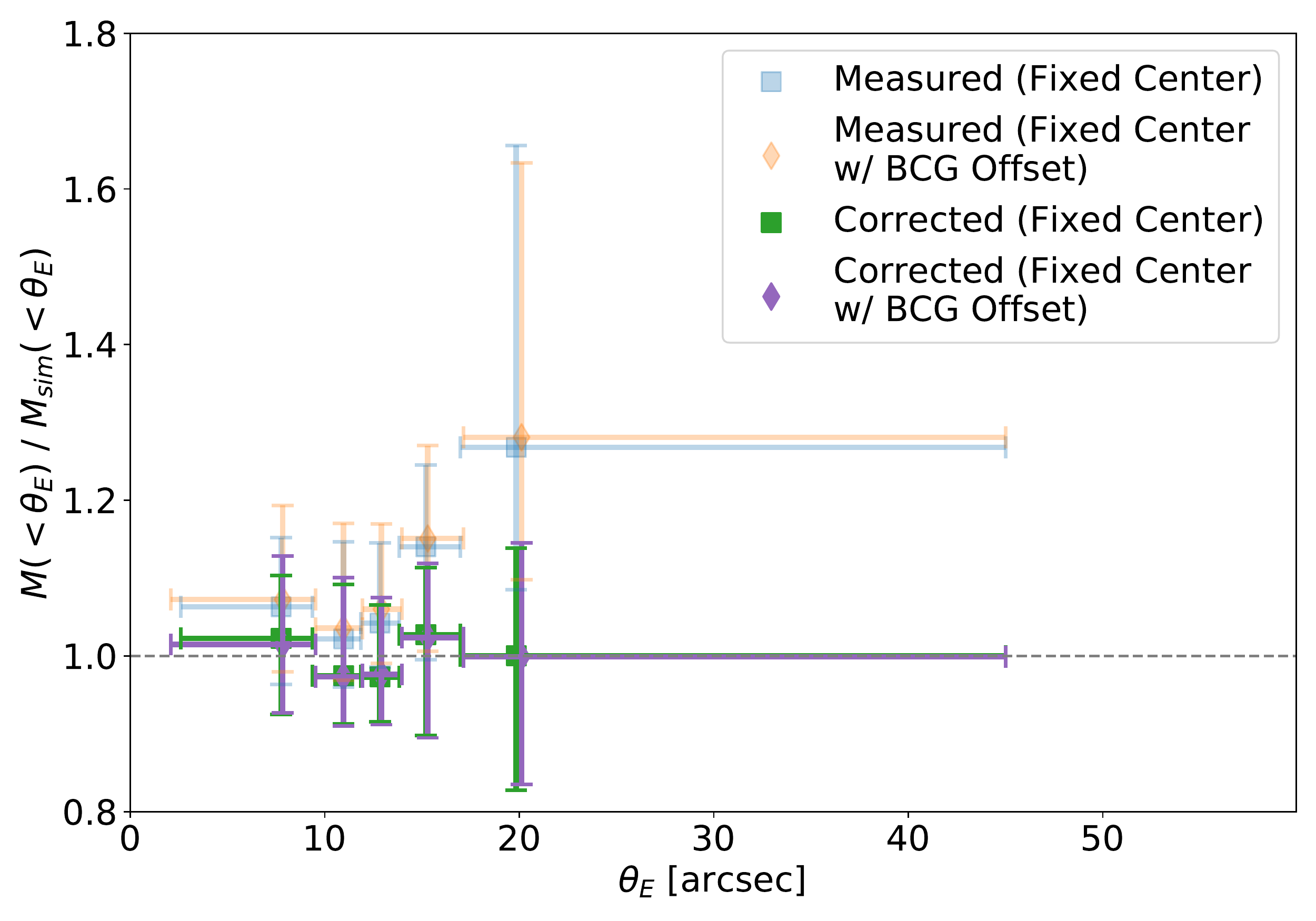}
\caption{\textsc{\textbf{Empirically Corrected Mass Ratio $\ML / \MT$ Binned by $\theta_E$.}} The blue and orange are from the analysis in \autoref{fig:er_binned_analysis}, while the green and magenta represent the empirically corrected values, using \autoref{eq:un_bias_ml}. The symbols and error bars are the same as \autoref{fig:er_binned_analysis}. We find that using the empirical correction un-biases and reduces the scatter of $\ML$.}
\label{fig:ec_er_binned_analysis}
\end{figure}

We plot in \autoref{fig:ec_er_binned_analysis} the empirically corrected values of $\ML$ and show the results from \autoref{fig:er_binned_analysis} for reference. With the mass enclosed by the Einstein radius corrected using the empirical correction, the overall scatter (half of the difference between the 84th and the 16th percentile of the distribution) reduces to  \FXECScatter\ (\FXBCGECScatter) and the bias to \FXECBias\ (\FXBCGECBias) for the fixed center (fixed center with BCG offset). 

\begin{figure*}
\center
\includegraphics[width=1.0\textwidth]{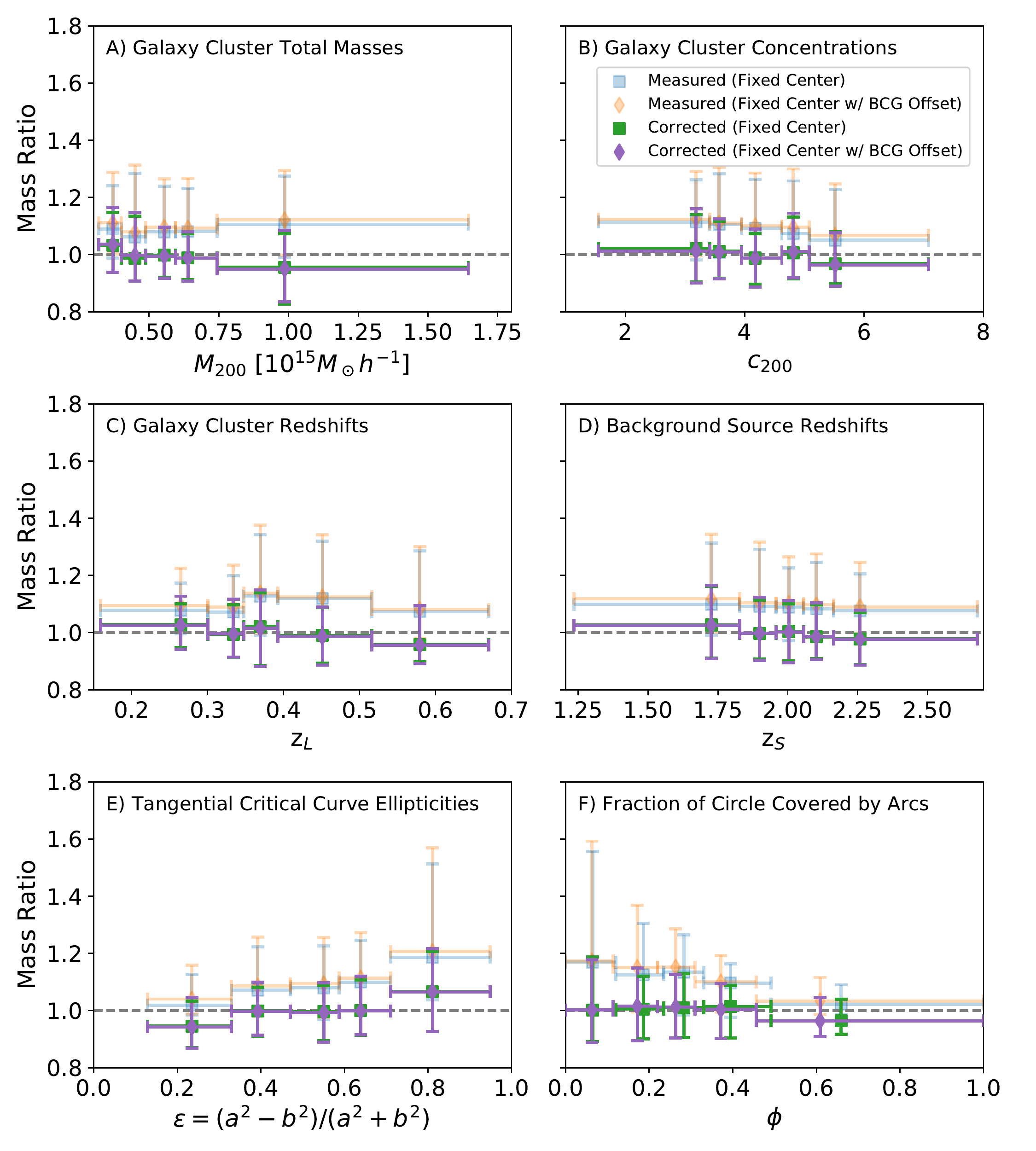}
\caption{\textsc{\textbf{Empirically-Corrected Inferred Mass Binned by Galaxy Cluster Properties, Background Source, and Lensing Geometry.}} Same as \autoref{fig:binned_analysis}, but using \autoref{eq:un_bias_ml} to empirically correct the mass estimates. The blue and orange points are from the analysis in \autoref{fig:binned_analysis}, while the green and purple represent the empirically corrected values. We find overall that using the empirical correction un-biases the results and reduces the scatter of $\ML$. The empirical correction does not introduce significant correlation with total cluster mass, concentration, or redshifts. It does not eliminate the trend due to deviation from circular symmetry, as can be seen in Panel E.}
\label{fig:ec_binned_analysis}
\end{figure*}

We then perform similar analyses as those in \S \ref{sec:analysis}. We explore the systematics in the mass enclosed by the Einstein radius when the empirical correction is applied, and plot the results in \autoref{fig:ec_binned_analysis}. The blue and orange are the same from \autoref{fig:binned_analysis} and are plotted for reference, while the green and red indicate the empirically corrected values. 

 We observe in \autoref{fig:ec_binned_analysis} that overall the measurement of the mass enclosed by the Einstein radius becomes un-biased. The scatter of $\ML$ is reduced in all the bins when compared to the analysis without empirical correction for the total mass, concentration, lens redshift, and background redshift. Using the empirical correction reduces the scatter in the highest-scatter bins, i.e., at high and low Einstein radius, small arc fraction, and large ellipticity of the tangential critical curve.


\section{Conclusions} 
\label{sec:conclusion}

With current and future large surveys discovering tens of thousands of clusters and groups, with thousands expected to show strong lensing features, an efficient method to estimate the masses at the cores of these systems is necessary. The mass enclosed by the Einstein radius is a quick zeroth-order estimate. Studies that use this method quote an uncertainty of $\sim 30\%$ (e.g., \citealt{Bartelmann:96, Schneider:06b}),  although this uncertainty has not been thoroughly quantified. In this work, we conduct a detailed analysis of the efficacy of the mass enclosed by the Einstein radius as core mass estimator, using the Outer Rim cosmological simulation. When measuring the Einstein radius, we explore three centering assumptions: fixed center, free center, and a observationally-motivated centering that mimics fixing the center to the BCG. We measure the scatter and bias of $\ML$, identify sources of systematic errors, and explore possible indicators available from imaging data at the cores of galaxy clusters. The results of our work are summarized below:

\begin{itemize}
    \item In the fixed center approach, the center of the circle is fixed to the highest surface density point and a circle is fitted to the tangential arcs. The statistical uncertainty in the measured Einstein radius is small (see \autoref{fig:ang_r_unc_dist}). We measure an overall scatter of \FXAllScatter\ with a bias of \FXAllBias\ in the mass enclosed by the Einstein radius with no correction applied.
    \item In the free center approach, the center of the circle is a free parameter in the fit. The statistical uncertainty of the Einstein radii fitted with the method is $20$ times higher than that of fixed center and the fixed center with BCG offset (see \autoref{fig:ang_r_unc_dist}). With this method, the overall scatter is \FRAllScatter\ with a bias of \FRAllBias\ in the mass enclosed by the Einstein radius with no correction applied. We do not recommend the use of the free center  method to measure the mass enclosed by the Einstein radius due to the large scatter in the mass measurement, high uncertainty in the Einstein radius, and restriction of a minimum of 3 or more identified tangential arcs.
    \item With the intention to apply this to observational data, we investigate the effect of using the BCG as the fixed center. We move the fixed center from the point of highest density by a random offset, following the log-normal distribution ($\mu = 6.1 \pm 0.7$ kpc) of BCG offsets found by \citet{Harvey:19}. This offset increases the scatter to \FXAllBCGScatter, and the bias to \FXAllBCGBias\ in the mass enclosed by the Einstein radius when compared to the fixed center method.
    \item We find that the scatter and bias of $\ML$ with respect to $\MT$ does not depend on the total cluster mass, concentration, lens redshift, or source redshift (\autoref{fig:binned_analysis}).
    \item We explore how the deviation from circular symmetry affects the measurement of $\ML$. The tangential critical curve ellipticity ($\epsilon$) stems from the deviation from spherical symmetry of the projected mass distribution at the core of the cluster. We find that the bias and scatter correlate with $\epsilon$ (\autoref{fig:binned_analysis}), where larger deviations from circular symmetry lead to a larger bias and scatter of $\ML$ when compared to $\MT$.
    \item The fraction of the circle covered by arcs of a single lensed source ($\phi$), can be directly accessed from the imaging data. This observable correlates strongly with the scatter and bias, with both scatter and bias decreasing with an increasing fractional coverage by the arcs (\autoref{fig:binned_analysis}). $\phi$ can be used as an observational indicator to estimate the field-specific scatter and bias of $\ML$ (\autoref{table:frac_arc_binned}).
    \item Other possible sources of systematic errors exits. While the Outer Rim simulation has a large volume and high mass resolution needed for this work, we are limited by the lack of baryonic information in the simulation and missing the structure along the line-of-sight in the simulated ray-traced images.
    For example, the structure along the line-of-sight, particularly in the case of low mass systems, will have an effect on this measurement \citep{Bayliss:14,Li:19}. We leave this investigation for future work.
    \item We evaluated the case when the background source redshift measurement is not available, using instead the distribution of the background source redshifts. While an accurate source redshift is critical for several lensing applications (e.g., magnifications, time delays, mass distribution) for the relatively well-constrained enclosed core mass, the scatter introduced by the uncertainty in the background source redshift is negligible compared to that of other systematics (\autoref{fig:nozs_binned_analysis}), if the underlying source redshifts distribution can be accurately estimated. In addition the dependence on the $\zs$ is predictable and matches our expectations, \S \ref{sec:no_zs} and \autoref{fig:nozs_binned_analysis}.
    \item We derive an empirical correction to un-bias and reduce the scatter of the measurement of $\ML$ using a quadratic equation fitted to the mass ratio ($\ML / \MT$) with respect to the Einstein radius. The scatter of the empirically corrected masses enclosed by the Einstein radius reduces to \FXECScatter\ and \FXBCGECScatter\ respectively for fixed center and fixed center with a BCG offset. The empirical correction does not introduce correlation between the inferred mass and other cluster or background source properties, which is important for application of this method in measuring cluster properties such as the concentration-mass relation as a function of redshift.
\end{itemize}

\subsection{Application}

In this section we provide a recipe for applying the results of this work to observational data, to statistically correct the bias in $\ML$ and estimate its uncertainty.

We note that a more accurate estimate of the field-specific uncertainty can be achieved by using the fraction of the Einstein circle covered by arcs as an indicator of deviation from circular symmetry. We provide instructions for both choices.

1) Starting with a cluster lens field in which lensing evidence has been detected, identify all the secure multiple images (arcs) of the lensed source. Each lensed image should be classified as either tangential or radial. Only the tangential arcs are used to estimate $\ML$. 

2) Measure the exact coordinates of a morphological feature (e.g., a bright emission clump) that repeats in each of the arcs.

3) Fit a circle to the list of coordinates. If the cluster has a distinct BCG, we recommend fixing the center of the fitted circle to the position of the BCG.
The radius of the fitted circle defines $\theta_E$.

4) Measure $\phi$, the fraction of the circle covered by the arcs of a single lensed source, by summing the angles subtended by the extent of the arcs that overlap with the Einstein circle, and dividing the sum by $360^{\circ}$. An example of three cases of different $\phi$ values is shown in \autoref{fig:frac_arc_example}.

5) Calculate $\ML$, the projected mass density enclosed in $\theta_E$, by evaluating \autoref{eq:s_crit} and \autoref{eq:m_er} for the cluster and source redshifts, and the measured $\theta_E$. 

If the spectroscopic redshift of the source is unknown, it can be approximated from photometric redshifts or a probability distribution function of source redshifts. we find that for the purpose of a statistical measurement of the enclosed mass, the increase in uncertainty due to a small error in the source redshift is negligible compared to other sources of uncertainty. 

6) Evaluate whether an empirical correction is beneficial:
If $\phi \gtrsim 0.5$ (i.e., the arcs of an individual lensed source cover at least half of the Einstein circle), the measured $\ML$ is fairly unbiased and an empirical correction is not necessary. In all other cases, or if the choice is to not use $\phi$ as an indicator, proceed to apply the empirical correction as follows.

7) Calculate $f(\theta_E)$, the empirical correction factor, by evaluating \autoref{eq:quad_eq} for $\theta_E$ (see \autoref{table:emp_corr_models} for coefficient values). We recommend using the fixed circle with BCG offset method. For Einstein radii in the range of $\theta_E < 30\farcs0$, we recommend using the quadratic fit.
Apply the correction to the measured $\ML$ using \autoref{eq:un_bias_ml}.

8)Determine the uncertainty. The field-specific uncertainty decreases as the fraction of the Einstein circle covered by arcs ($\phi$) increases. The numerical values of the scatter as well as the 16th and 84th percentiles (lower and upper limit) for five $\phi$ bins are tabulated in \autoref{table:frac_arc_binned} in \autoref{appsec:frac_arc_analysis}.
If the $\phi$ estimator is not used, one can assume an overall uncertainty in the corrected $\ML$ of \FXECScatter\ (\FXBCGECScatter) for the fixed center (fixed center with BCG offset).  \\

With the characterization of the mass enclosed by the Einstein radius presented in this work --- including the application of indicators of the scatter and bias --- measuring the mass at the cores of strong lensing galaxy clusters can be performed in large samples in a very efficient manner. The estimation of the mass at the core can be used to determine the mass distribution profile of the galaxy cluster, the concentration parameter (when combined with a mass estimate at larger radius), and provide information about the baryonic and dark matter properties at the core of galaxy clusters.


\section*{Acknowledgements}

The authors would like to thank the anonymous referee for insightful suggestions that improved this manuscript. This material is based upon work supported by the National Science Foundation Graduate Research Fellowship Program under Grant No. DGE 1256260. Work at Argonne National  Lab is supported by UChicago Argonne LLC, Operator of Argonne National Laboratory. Argonne National Lab, a U.S. Department of Energy Office of Science Laboratory, is operated by UChicago Argonne LLC under contract no. DE-AC02-06CH11357. This research used resources of the Argonne Leadership Computing Facility, which is a DOE Office of Science User Facility supported under Contract DE-AC02-06CH11357.


\clearpage

\appendix

\section{Uncertainty dependence on the fraction of circle covered by arcs}
\label{appsec:frac_arc_analysis}

In this appendix we give numerical values of the field-specific uncertainty, which depends on the deviation from circular symmetry, as indicated by the fraction of the circle covered by arcs ($\phi$). For the statistics used in our analysis see \S\ref{subsec:stats}. The scatter is defined as half the difference between the 84th percentile (upper) and the 16th percentile (lower) of the distribution and we compute the bias using the median of the distribution. For convenience, we tabulate the numerical values that are plotted in \autoref{fig:ec_binned_analysis} in  \autoref{table:frac_arc_binned}.

\capstartfalse
\begin{deluxetable*}{cccc}[h]
\tablecolumns{4}
\tablewidth{2\columnwidth}
\tablecaption{Bias and uncertainty in $\ML$ as a function of $\phi$}
\tablehead{ & & \multicolumn{2}{c}{$\ML$/$\MT$ } \\
& $\phi$ bin & Measured  & Corrected \\
& \colhead{median [min --- max]} & \colhead{median [lower --- upper]}  & \colhead{median [lower --- upper]} }
\startdata
 Fixed Center & 0.06 [0.00 --- 0.12] & 1.17 [0.99 --- 1.56] & 1.00 [0.89 --- 1.19] \\
 & 0.19 [0.12 --- 0.23] & 1.13 [0.99 --- 1.30] & 1.01 [0.90 --- 1.12] \\
 & 0.28 [0.23 --- 0.33] & 1.13 [0.99 --- 1.26] & 1.01 [0.91 --- 1.13] \\
 & 0.39 [0.33 --- 0.49] & 1.10 [0.98 --- 1.16] & 1.01 [0.90 --- 1.09] \\
 & 0.66 [0.49 --- 1.00] & 1.02 [0.99 --- 1.09] & 0.96 [0.92 --- 1.04] \\
 \hline
 Fixed Center & 0.06 [0.00 --- 0.11] & 1.17 [1.00 --- 1.59] & 1.00 [0.89 --- 1.18] \\
 w/ BCG Offset & 0.17 [0.11 --- 0.22] & 1.15 [1.00 --- 1.37] & 1.02 [0.89 --- 1.15] \\
 & 0.26 [0.22 --- 0.31] & 1.15 [1.00 --- 1.29] & 1.01 [0.90 --- 1.13] \\
 & 0.37 [0.31 --- 0.46] & 1.10 [0.99 --- 1.19] & 1.00 [0.90 --- 1.09] \\
 & 0.61 [0.46 --- 1.00] & 1.03 [0.99 --- 1.12] & 0.96 [0.91 --- 1.05] \\

\enddata
\tablenotetext{}{A quantitative form of the information displayed in Panel~F of \autoref{fig:binned_analysis} and \autoref{fig:ec_binned_analysis}. The median and boundaries of the bins of $\phi$ are tabulated in the first column; the next columns tabulate the median, lower 16th percentile, and the upper 84th percentile  of $\ML$/$\MT$, for the measured results (\autoref{fig:binned_analysis}) and corrected results (\autoref{fig:ec_binned_analysis}). }
\label{table:frac_arc_binned}
\end{deluxetable*}
\capstarttrue

\bibliographystyle{yahapj}
\bibliography{bibfile}

\end{document}